# Cross-View and Cross-Time Tracking of Crack Evolution in Concrete

Pengru Deng [a], Jiapeng Yao [a, *], Su Wang [c], Xuhui He [a], Zhao Wang [a], Takanori Emaru [d], Takashi Matsumoto[d], and Xinrun Li [b, **]

[a] School of Civil Engineering, Central South University, Changsha, China

[b] School of Computing, Newcastle University, Tyne and Wear, United Kingdom

[c] Nvidia, Shanghai, China

[d] Faculty of Engineering, Hokkaido University, Sapporo 060-8628, Japan

*Corresponding authors, E-mail: jiapengyao@csu.edu.cn, xinrun.li@ncl.ac.uk.

**Abstract:** Crack propagation provides richer information than static crack observations for evaluating structural conditions of concrete infrastructure and planning maintenance strategies. Unlike static crack detection, which can now be largely automated using computer vision, vision-based crack tracking is a coupled cross-view and cross-time problem: the same physical crack must retain its identity as camera viewpoints change, while newly developed growth must be distinguished from existing crack segments in a unified reference view. Existing segmentation and 3D reconstruction methods can recover crack masks or scene geometry, but they do not preserve crack identity across views or quantify temporal extension. To address this gap, this study formulates crack growth monitoring as a cross-view and cross-time crack-instance tracking task and proposes a geometry-assisted, morphology-aware framework. The framework first uses image-based geometric registration to align observations from different viewpoints into a shared structural reference, then applies morphology-aware instantiation to assign physically interpretable crack identities, and finally performs temporal mapping in the unified view to distinguish newly developed growth from existing crack segments. To enable evaluation, a temporally annotated dataset was established and used, covering three representative cracking scenarios: single dominant crack propagation, mixed-mode branching, and multiple-crack interaction, with pixel-level masks and cross-time instance labels. Experimental results show that the proposed framework can establish loose cross-view identity correspondence under changing viewpoints, while strict spatial overlap and growth localization remain limited by early micro-crack segmentability and projection drift, highlighting directions for future crack-monitoring research.

## 1. Introduction

Crack propagation is a direct indicator of damage evolution in concrete structures and provides important evidence for structural condition assessment [1, 2]. Concrete deterioration is essentially a progressive process involving micro-crack initiation, propagation, and eventual instability under mechanical loading, environmental exposure, and aging. During this process, cracks may remain inactive or gradually become active, with their evolution typically involving stable propagation followed by accelerated growth before severe deterioration or failure. Therefore, the key concern in structural inspection is not only whether cracks exist, but whether they continue to propagate and how rapidly they develop. The propagation rate of active cracks provides important information for evaluating structural safety and determining appropriate maintenance strategies. This necessity for such monitoring is also recognized in major bridge inspection standards, including the Chinese *Technical Specifications for Field Inspections of Existing Highway Bridges* [3], the Japanese *Guideline for Periodic Road Bridge Inspection* [4], and the *American Association of State Highway and Transportation Officials (AASHTO) Manual for Bridge Evaluation* [5]. These standards call for active cracks to be monitored continuously or periodically according to their growth rates and activity levels.

In practice, crack propagation is commonly monitored using embedded sensors or contact-based nondestructive testing (NDT) methods in concrete infrastructures. Although these methods can provide useful measurements, they are often limited by installation cost, accessibility, sensing coverage, and dependence on predefined monitoring locations. Similar limitations also exist in laboratory investigations, where crack paths are often manually traced at different loading stages, making continuous and high-frequency crack-growth observation difficult. Vision-based methods offer a flexible, non-contact alternative and have been widely applied to crack detection and measurement [6, 7]. Deep learning-based segmentation models can effectively extract thin, irregular, and low-contrast cracks from complex backgrounds, providing pixel-level crack masks for quantitative measurement [8–10]. However, these methods mainly address single-image crack perception and cannot determine whether cracks observed at different times or viewpoints correspond to the same physical crack. A crack mask obtained from an individual image only describes the spatial appearance at one observation moment. It cannot answer two fundamental questions in crack-growth monitoring: whether the observed crack is the same crack tracked previously and how much new crack extension has occurred. Therefore, moving from static crack recognition toward crack-growth monitoring requires maintaining the identity of each physical crack across observations rather than interpreting each crack mask independently. Accordingly, crack-growth monitoring from images is formulated as a cross-view and cross-time crack-instance tracking task.

Beyond crack segmentation in individual images, crack-instance tracking seeks to maintain the identity of each physical crack across time and quantify its spatiotemporal evolution under varying observation conditions. Cross-view and cross-time crack-instance tracking comprises four coupled sub-problems: (1) cross-view geometric alignment, which establishes correspondence between images acquired from different viewpoints; (2) crack-instance identification, which converts pixel-level crack masks into individual crack entities corresponding to physical cracks; (3) crack-instance association, which determines whether crack observations from different times belong to the same physical crack; and (4) growth separation and quantification, which distinguishes inherited crack regions from newly developed extensions and measures crack growth. These sub-problems must be addressed jointly to obtain consistent crack-growth histories under changing viewpoints.

Existing studies have addressed several components of this task, but these components have largely been studied separately. For cross-view alignment, image stitching and feature-based registration methods have been developed to enlarge observation coverage and improve spatial consistency. Panoramic reconstruction supports large-area crack mapping [11, 12], while dense and learning-based feature matching improves correspondence estimation between overlapping crack images [13–15]. Crack-specific registration approaches further enable multi-view crack detection and width measurement [16–22]. However, these methods mainly address single-time spatial alignment and do not consider crack evolution.

For temporal crack tracking, most previous studies have maintained image correspondence by keeping the camera fixed or constraining its motion. Kong et al. [23] used shape-context matching to retain crack labels across loading stages recorded from a fixed camera position. Xu et al. [24] monitored consecutive video frames acquired by either a stationary camera or a camera moving along a fixed guide rail. Chen et al. [25] quantified crack development using a pre-calibrated camera that continuously observed the same beam region, thereby maintaining a stable viewpoint during loading. These studies demonstrated the feasibility of temporal monitoring, but their comparisons depended on either a fixed viewpoint or a predefined camera trajectory. Gao et al. [26], in contrast, considered camera-pose variation by establishing a structural coordinate system from a salient positioning weld and its endpoint. This strategy permits camera repositioning, but it relies on a structure-specific landmark that is often unavailable on general concrete surfaces. More general crack monitoring therefore requires geometric correspondence to be recovered directly from overlapping images when the camera pose is freely changed between observations, without relying on a predefined camera trajectory or a structure-specific landmark.

Recent advances in image matching and 3D reconstruction have made it possible to recover geometric relationships directly from overlapping images. By establishing spatial correspondence between views, these advances make it possible to compare observations of the same physical crack after the camera has been freely repositioned. Modern image matching methods, such as SuperGlue based on graph neural networks [13] and detector-free dense correspondence models such as ASpanFormer [14], enable robust feature-level or pixel-level matching under significant viewpoint variations. Meanwhile, Structure-from-Motion (SfM), image-based 3D reconstruction, and emerging 3D-aware correspondence models such as DUSt3R and MASt3R can recover camera poses and establish a shared geometric reference for observations captured from different viewpoints [16–19, 27–29]. These developments make it possible to map time-lapse images into a unified coordinate system, where crack evolution can be compared consistently. However, geometric registration alone only solves the spatial correspondence problem. It does not determine crack identity, distinguish newly formed growth from existing crack regions, or provide crack-level temporal evolution.

Therefore, complete crack-instance tracking requires integrating geometric alignment with crack instance representation, association, and growth measurement. It also requires a dedicated evaluation protocol tailored to crack-like structures that are thin, deformable, and capable of elongating and branching over time. Inspired by these developments, this study proposes a unified framework for cross-view and cross-time crack-instance tracking in concrete structures. The framework first recovers structural geometry from the available overlapping views. For each successive observation pair, the recovered geometry is then used to map the earlier image into the coordinate frame of the later image. The aligned crack masks are subsequently organized into crack instances, whose identities are associated across successive observations. Finally, inherited crack regions are separated from newly appeared growth for quantitative analysis. The overall workflow is illustrated in Fig. 1. The main contributions are summarized as follows:

1. This study formulates crack-instance tracking for visual crack-growth monitoring under changing viewpoints, enabling cross-time identity association and growth detection using overlapping images acquired with freely chosen imaging devices from arbitrary viewpoints.

2. A geometry-assisted and morphology-aware framework is developed to integrate image registration, crack instantiation, and unified-view temporal mapping, converting binary crack masks into physically interpretable instances, propagating crack identities, and separating inherited crack support from newly formed growth.

3. A crack-growth tracking evaluation protocol is proposed to assess identity consistency, growth localization, and propagated-length quantification across viewpoints and time.

4. A temporally annotated crack-image evaluation dataset is established, with pixel-level masks and cross-time instance labels to support joint evaluation of identity preservation and growth attribution across representative cracking patterns.

# 2. Methodology

## 2.1 Framework of the proposed method

The proposed approach is designed to enable accurate and interpretable analysis of crack extension by integrating image crack segmentation, cross-view geometric registration, crack instance identification, and temporal extension tracking. Instead of treating cracks as isolated image patterns, it represents them as physical structural entities observed across stages within a shared geometric reference space. As illustrated in Fig. 1, the framework consists of four modules: (1) geometric structure estimation, (2) image crack segmentation, (3) crack instance identification, and (4) crack numbering and extension tracking.

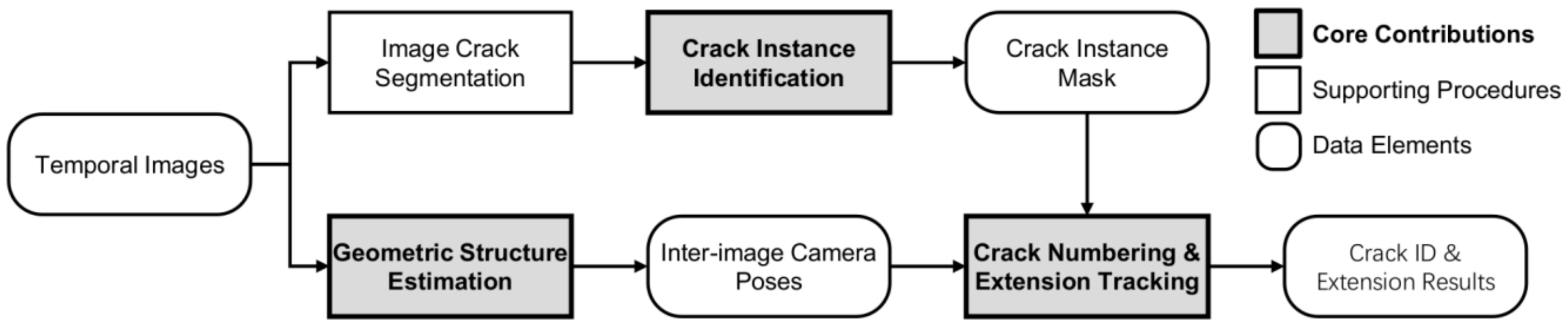

**Fig. 1.** Overall workflow of the proposed crack instance matching and extension-tracking framework.

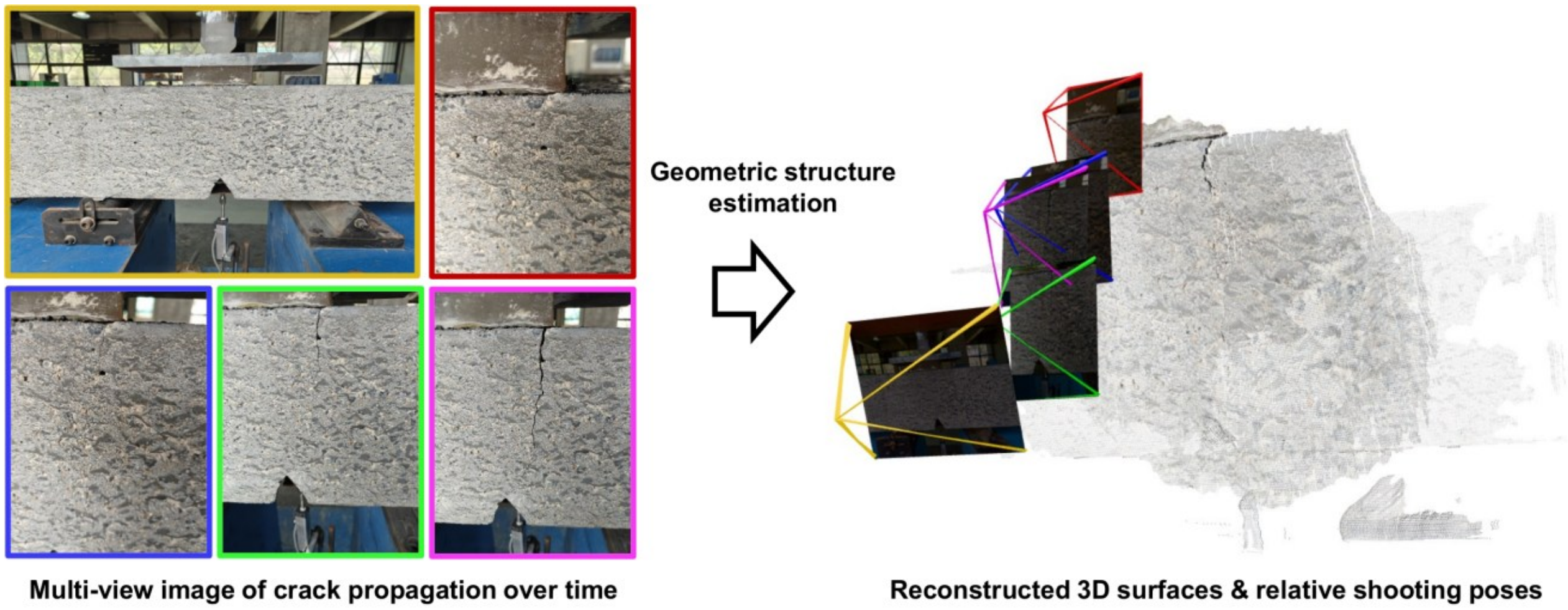

**Fig. 2.** Geometric structure estimation from temporal crack images. The module estimates dense scene geometry and relative camera poses from multi-view crack images to support unified-view crack registration.

### 2.2 Geometric structure estimation

Viewpoint variation disrupts the pixel-wise correspondence between observations of the same physical crack. Even when the crack remains unchanged, camera motion changes its apparent position, orientation, scale, and local shape in the image plane. This effect is critical for crack tracking because cracks are narrow and often fragmented. A small registration error can substantially reduce mask overlap and may be mistaken for temporal growth. The geometric module therefore maps observations from different viewpoints into a shared structural reference, as shown in Fig. 2. The aligned crack masks then provide the basis for crack-instance association and growth separation.

Cross-view mappings can be estimated using classical SfM or learned local-feature matching methods [13, 15, 29]. Surface cracks develop on three-dimensional structural members, so their apparent displacement across views is governed by both camera motion and the geometry of the structural surface. Although a small local surface patch may be approximated as planar, an inspection image may span multiple faces or contain depth variation, locally non-planar geometry, and occlusion. Such conditions are particularly pronounced at bridge piers and structural corners. A single global homography may therefore be unable to describe the pose-induced displacement over the entire crack region. The registration front end should consequently provide dense local support around narrow crack segments while accounting for the three-dimensional geometry of the supporting structural surface.

These requirements motivate the use of MASt3R. MASt3R jointly predicts dense image descriptors and 3D point maps and establishes correspondences through reciprocal matching [27, 28]. The dense descriptors reduce reliance on isolated keypoints near narrow or fragmented crack regions, while the 3D point maps provide spatially varying mappings without imposing a single global planar model. In the current implementation, the resulting point maps, valid-pixel masks, camera poses, and focal lengths are used to construct source-to-target mappings for downstream crack tracking. The pretrained MASt3R backbone is used without retraining or modification. The registration front end is modular, allowing MASt3R to be replaced by any method that provides dense, geometry-aware correspondences without changing the downstream tracking pipeline.

Within MASt3R, dense correspondences are established by reciprocal nearest-neighbor matching between the descriptor fields of the two views. Efficient reciprocal matching and local refinement avoid exhaustive comparisons over all pixel pairs [28]. These operations produce the geometric evidence used to instantiate the registration-induced mapping defined below.

For notational convenience, consider one overlapping view pair from the multi-view sequence, denoted as $I_1$ and $I_2$. The geometric front end is abstracted as a registration-induced mapping $T_{1\to2}$ from source-view image coordinates to the target or reference view:

$$\boldsymbol{x}_2 = T_{1\to2}\left(\boldsymbol{x}_1;\Theta_{12}\right) \tag{1}$$

where $\boldsymbol{x}_1$ is a source-view image coordinate, $\boldsymbol{x}_2$ is its mapped position in the target or reference view, and $\Theta_{12}$ denotes the estimated geometric relationship between the two views. Depending on the selected backend, $\Theta_{12}$ may be a homography, SfM-derived camera poses, or dense correspondence with local 3D support. The crack-tracking module only consumes the resulting mapping and confidence information, so the same formulation can support different pose-recovery methods without changing the downstream crack-instance and growth-tracking logic.

### 2.3 Image crack segmentation

Crack segmentation provides the visible crack support from which tracking starts, but it is not treated as the main contribution of this study. The segmentation front end adopts the DeepLabv3+ checkpoint developed in the previous study [30], without retraining or fine-tuning on any image from the temporal scenes evaluated here. As shown in Fig. 1, its predicted masks are used directly as inputs to crack-instance identification and temporal tracking. This dependence on predicted masks is important for interpretation: weak segmentation of early micro-cracks can propagate into fragmented instances, reduced cross-view overlap, and delayed growth localization in later evaluation.

### 2.4 Crack instance identification

The binary crack mask identifies crack pixels but does not distinguish individual physical cracks. Two complementary errors must therefore be addressed. A connected foreground component may contain branches belonging to different crack systems at a crossing, whereas one physical crack may appear as several disconnected fragments because of weak contrast, surface texture, or local occlusion. A morphology-aware, split-and-merge strategy is therefore adopted. Each 8-connected foreground component $\boldsymbol{\Gamma}_i^{(0)}$ is first skeletonized and decomposed into primitive branches at endpoints and junctions. Branches whose dominant directions, estimated over their full skeleton extents, differ by a sufficiently large angle are assigned a hard-separation constraint according to the rule defined below. The remaining disconnected fragments are subsequently merged when they satisfy the morphology-based continuity criteria. No component-area filtering is applied, and all predicted foreground pixels are retained at this instantiation stage.

To obtain primitive crack branches and their representative endpoints, each initial 8-connected component $\boldsymbol{\Gamma}_i^{(0)}$ is reduced to a one-pixel-wide skeleton using the Zhang-Suen parallel thinning algorithm [31]. For a candidate foreground pixel with binary value $p_1 = 1$, let $p_2, \ldots, p_9 \in \{0,1\}$ denote the binary values of its eight neighboring pixels, ordered clockwise from the north, where 1 and 0 represent foreground and background, respectively. The neighborhood properties are defined as:

$$B(p_1) = \sum_{k=2}^{9} p_k, \qquad A(p_1) = \sum_{k=2}^{9} (1 - p_k) p_{k+1}, \qquad p_{10} = p_2. \tag{2}$$

where $B(p_1)$ counts the foreground neighbors and $A(p_1)$ counts the background-to-foreground transitions in the circular sequence $p_2, \ldots, p_9, p_2$. The index $k$ enumerates the eight neighbors, and $p_{10} = p_2$ closes the sequence. The first and second thinning sub-iterations use the following deletion conditions:

$$\begin{aligned} \mathcal{C}_1(p_1):&\quad 2 \le B(p_1) \le 6, \quad A(p_1) = 1, \quad p_2 p_4 p_6 = 0, \quad p_4 p_6 p_8 = 0, \\ \mathcal{C}_2(p_1):&\quad 2 \le B(p_1) \le 6, \quad A(p_1) = 1, \quad p_2 p_4 p_8 = 0, \quad p_2 p_6 p_8 = 0, \end{aligned} \tag{3}$$

where $\mathcal{C}_1(p_1)$ and $\mathcal{C}_2(p_1)$ denote the deletion conditions for the first and second thinning sub-iterations, respectively. The neighborhood quantities $B(p_1)$ and $A(p_1)$ are defined in Eq. (2). A candidate foreground pixel is removed only when all conditions in the corresponding row are satisfied. Alternating these two sub-iterations until convergence yields the skeleton:

$$\boldsymbol{S}_i = \mathrm{Skel}\left(\boldsymbol{\Gamma}_i^{(0)}\right), \tag{4}$$

where $\mathrm{Skel}(\cdot)$ denotes the Zhang-Suen thinning operation governed by Eq. (3), $\boldsymbol{\Gamma}_i^{(0)}$ is the binary mask of the *i*-th initial 8-connected component, and $\boldsymbol{S}_i$ is the resulting one-pixel-wide skeleton. Eligible pixels are removed in parallel within each sub-iteration, and the neighborhood conditions are reevaluated on the updated mask before the next sub-iteration. The process stops when neither sub-iteration removes any further pixels. Skeleton pixels are classified as endpoints, ordinary pixels, or junctions according to their 3 × 3 connectivity. Paths traced between endpoints and junctions define primitive crack branches. For notational simplicity, the resulting branches are reindexed as $\boldsymbol{\Gamma}_i$ in the continuity analysis below. The terminal pixels of each primitive branch are used as representative endpoints; when fewer than two reliable endpoints are available, the extreme points along its principal axis are used instead.

Let $\boldsymbol{P}_i \in \mathbf{R}^{N_i \times 2}$ denote the pixel-coordinate matrix of primitive branch $\boldsymbol{\Gamma}_i$, where its $n$-th row is $\boldsymbol{p}_{i,n}^{\mathsf{T}}$, and $\boldsymbol{p}_{i,n} = \left[ y_{i,n}, x_{i,n} \right]^{\mathsf{T}}$ denotes the image coordinate of the $n$-th foreground pixel. The centroid and principal direction of the branch are obtained by

$$\begin{gathered} \boldsymbol{c}_i = \frac{1}{N_i} \sum_{n=1}^{N_i} \boldsymbol{p}_{i,n}, \\ \tilde{\boldsymbol{p}}_{i,n} = \boldsymbol{p}_{i,n} - \boldsymbol{c}_i, \quad n = 1, \ldots, N_i \\ \tilde{\boldsymbol{P}}_i = \boldsymbol{U}_i \boldsymbol{\Sigma}_i \boldsymbol{V}_i^{\top}, \boldsymbol{u}_i = \boldsymbol{V}_i(:,1). \end{gathered} \tag{5}$$

where $N_i$ is the number of pixels in $\boldsymbol{\Gamma}_i$, and $\boldsymbol{c}_i$ is the centroid of the branch. The centered coordinate $\tilde{\boldsymbol{p}}_{i,n}$ is obtained by subtracting $\boldsymbol{c}_i$ from the coordinate of the $n$-th pixel. Stacking these centered coordinates row by row forms the centered coordinate matrix $\tilde{\boldsymbol{P}}_i$. In the singular value decomposition, $\boldsymbol{U}_i$ and $\boldsymbol{V}_i$ are the left- and right-singular-vector matrices, respectively, and $\boldsymbol{\Sigma}_i$ is the diagonal matrix of singular values. The unit vector $\boldsymbol{u}_i$, given by the first column of $\boldsymbol{V}_i$, represents the principal direction of $\boldsymbol{\Gamma}_i$. The principal-axis extent is calculated as the difference between the maximum and minimum projections of the component pixels onto $\boldsymbol{u}_i$. A direction estimate is considered reliable when the fragment contains at least four pixels and has a principal-axis extent of at least eight px. These limits determine whether directional information can be used in fragment matching. They are not used to remove small fragments.

Let $\boldsymbol{e}_i^{(a)}$ and $\boldsymbol{e}_j^{(b)}$, $a, b \in \{1, 2\}$, denote the two representative endpoints of $\boldsymbol{\Gamma}_i$ and $\boldsymbol{\Gamma}_j$, respectively. The minimum endpoint gap between two fragments is

$$d_{min}\left(i, j\right) = \min_{a,b \in \{1,2\}} \| \boldsymbol{e}_i^{(a)} - \boldsymbol{e}_j^{(b)} \|_2, \tag{6}$$

where a and b index the two endpoints of $\boldsymbol{\Gamma}_i$ and $\boldsymbol{\Gamma}_j$ respectively, and $\|\bullet\|_2$ denotes the Euclidean

norm. Among fragment pairs not subject to the hard separation rule, only those satisfying $d_{\min}(i,j) \le g_{\text{merge}}$ are considered as merge candidates, where $g_{\text{merge}}$ is the maximum candidate endpoint gap.

For fragments with reliable direction estimates, directional consistency is measured by

$$\theta_{ij} = \arccos\left(\left|\boldsymbol{u}_i^{\mathrm{T}}\boldsymbol{u}_j\right|\right), \tag{7}$$

where $\theta_{ij}$ is the acute angle between the principal axes of $\boldsymbol{\Gamma}_i$ and $\boldsymbol{\Gamma}_j$. The absolute value makes the comparison independent of the signs assigned to the two principal-direction vectors.

Because short discontinuities may occur along locally curved cracks, the allowable directional difference depends on the endpoint gap:

$$\theta_{\text{local}}(i,j) = \max\left[\theta_0 - \Delta\theta_{\max}, \theta_0 - \alpha \max\left(d_{\min}(i,j) - g_{\text{close}}, 0\right)\right], \tag{8}$$

where $\theta_0$ is the directional tolerance for short gaps, $\Delta\theta_{\max}$ is the maximum reduction in this tolerance, $\alpha$ is the distance-dependent decay coefficient, and $g_{\text{close}}$ is the endpoint-gap breakpoint beyond which the tolerance begins to decrease. This formulation accommodates local curvature across short mask interruptions while requiring stronger directional agreement between more widely separated fragments.

Directional agreement alone may incorrectly join nearby parallel cracks. The normal offset is therefore used to measure the sideways separation between two fragments. Before calculating this offset, the sign of $\boldsymbol{u}_j$ is aligned with that of $\boldsymbol{u}_i$. Let $\boldsymbol{n}_{ij}$ denote the unit normal to the normalized mean of the aligned directions. The normal offset is defined as

$$d_n(i,j) = \left|(\boldsymbol{c}_j - \boldsymbol{c}_i)^{\mathrm{T}}\boldsymbol{n}_{ij}\right|, \tag{9}$$

where $d_n(i,j)$ is the separation of the two fragment centroids along their common normal direction, and $\boldsymbol{n}_{ij}$ is the corresponding unit normal vector.

The allowable normal offset increases moderately with the endpoint gap:

$$d_{\mathrm{merge}}(i,j)=d_0+\lambda_n \min\left[d_{\min}(i,j),g_{\mathrm{merge}}\right]. \tag{10}$$

where $d_{\mathrm{merge}}(i,j)$ is the gap-dependent upper bound applied to the measured normal offset $d_n(i,j)$, $d_0$ is the base normal-offset tolerance for closely spaced fragments, and $\lambda_n$ controls how this tolerance increases with the endpoint gap. The term $\min\left[d_{\min}(i,j),g_{\mathrm{merge}}\right]$ caps this increase at the maximum candidate endpoint gap $g_{\mathrm{merge}}$.

For two fragments with reliable direction estimates that are not subject to the hard separation rule, the three core merge conditions are

$$\begin{gathered}\theta_{ij}\le\theta_{\mathrm{local}}(i,j),\quad d_n(i,j)\le d_{\mathrm{merge}}(i,j),\\ d_{\min}(i,j)\le g_{\mathrm{merge}}.\end{gathered} \tag{11}$$

The first condition requires compatible propagation directions, the second limits sideways separation, and the third prevents merging across an excessive endpoint gap. For larger endpoint gaps, the direction connecting the closest endpoints is also required to align with the principal axes of the fragments.

More restrictive distance and endpoint-axis alignment limits are applied when one or both direction estimates are unreliable. These limits distinguish the cases of two unreliable directions, direct attachment with one reliable direction, and axis-aligned attachment with one reliable direction. These rules allow small or poorly segmented fragments to be retained without relying on unstable directional estimates.

At crossing or large-turn intersections, two branches are topologically connected at a junction, so connectivity alone cannot separate them. To guide crack-instance separation, the angle between the full-branch dominant directions is defined as

$$\theta_{\mathrm{cross}}=\arccos(|\boldsymbol{t}_1^{\mathrm{T}}\boldsymbol{t}_2|), \tag{12}$$

where $\boldsymbol{t}_1$ and $\boldsymbol{t}_2$ are the unit dominant-direction vectors of the two branches, estimated from their full skeleton extents. Owing to the absolute value, $\theta_{\mathrm{cross}}$ lies in the range $[0^{\circ},90^{\circ}]$. When both branch-direction estimates are reliable and

$$\theta_{\text{cross}} \geq \theta_{\text{split}}, \qquad \theta_{\text{split}} = 45°, \tag{13}$$

the two branches are constrained to remain in different final instances. Physically, branches that intersect at large angles follow different local stress trajectories and belong to distinct crack systems, whereas a small intersection angle corresponds to a continuation or local branching of the same crack. The 45° threshold is consistent with the crack-direction zoning used in the nonlinear analysis of reinforced concrete: in the four-way fixed crack model, the angular domain is divided into 45° regions, each associated with one independent crack direction, and a crack that deviates beyond the region of an existing crack is treated as independent [32].

Finally, each initial fragment $\boldsymbol{\Gamma}_i$ is represented as a node in an undirected graph. An edge is added between two nodes when their closest representative-endpoint pair satisfies the merge conditions in Eq. (11), provided that the pair is not excluded by the crossing-angle separation rule in Eqs. (12)–(13). The connected groups are obtained using constrained union-find, which rejects any union that would place a hard-separated branch pair in the same group. All fragments in the same group receive the same instance label. This transitive grouping reconnects fragmented observations of the same crack while preserving every foreground pixel in the predicted binary mask. The complete procedure is illustrated in Fig. 3.

### 2.5 Crack numbering and crack-growth tracking

After geometric mapping and crack instantiation, the remaining issue is twofold and coupled. First, for each later observation, it must be decided whether it inherits the identity of an existing physical crack (identity continuity). Second, the newly appeared part within it must be attributed to that crack and measured as its growth (growth attribution). Mis-association directly corrupts the measured growth, so the two steps cannot be treated independently. Concretely, a crack segment observed at a later instant must either inherit the identity of a previous physical crack, be treated as a newly appeared crack, or be assigned as newly appeared growth of an existing crack. This module therefore defines how crack identities are propagated and how new support is measured in a unified reference view.

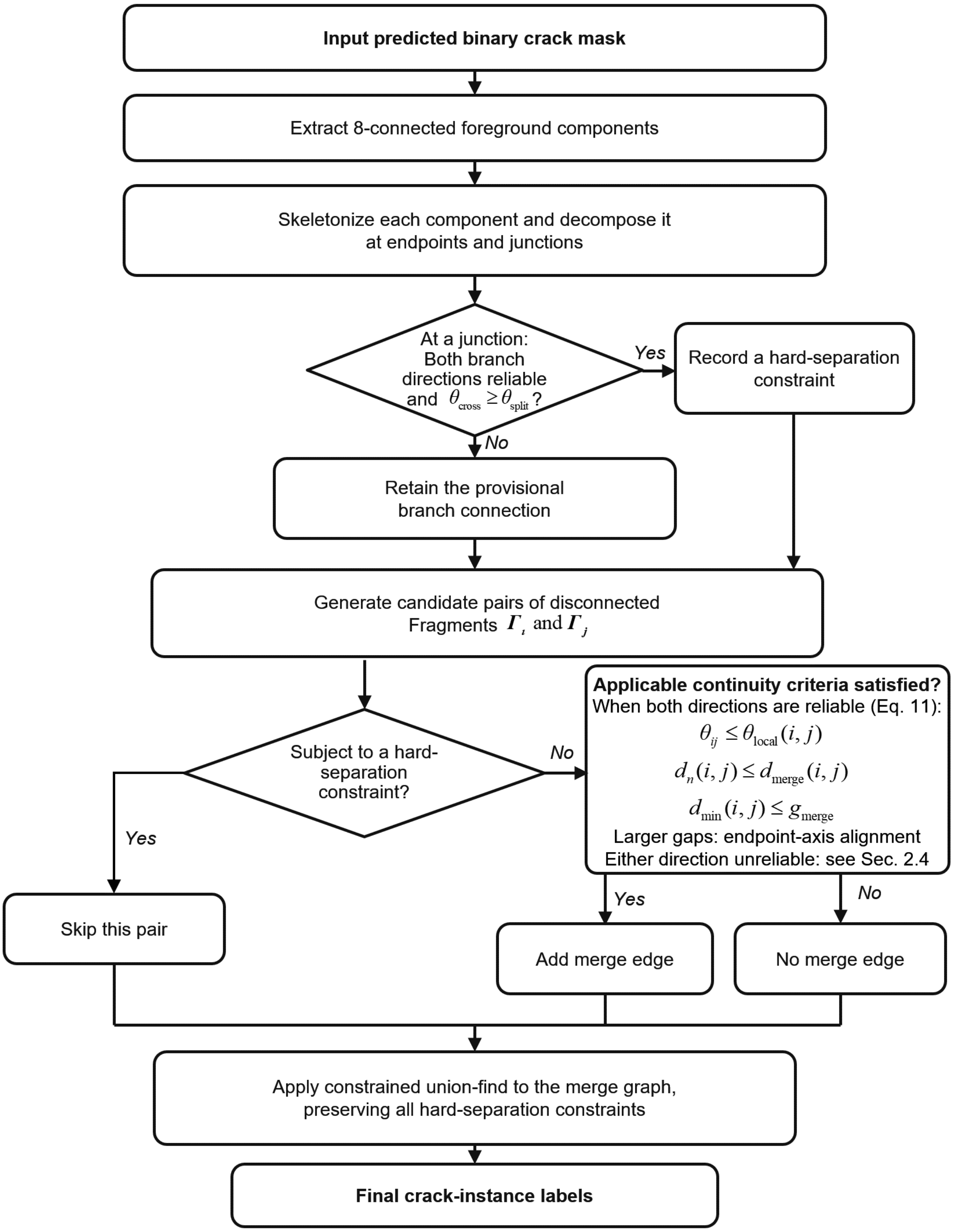

**Fig. 3.** Morphology-geometry decision process for crack-instance partition.

This section details how crack identities and growth are tracked across observations. As illustrated in Fig. 4, the module consists of three steps: (i) pose-guided projection and instance matching (Section 2.5.1), which projects instances into a common frame and establishes pairwise correspondence; (ii) crack numbering (Section 2.5.2), which propagates identities chronologically so that the same physical crack keeps one identifier; and (iii) pairwise growth extraction and length quantification (Section 2.5.3), which separates inherited skeleton support from newly appeared extension in the unified reference view. Note that two consecutive observations generally differ in both viewpoint and acquisition time; the registration and comparison below are designed precisely to decouple such observation-induced appearance changes from genuine structural growth. The novelty lies in treating identity propagation and growth attribution as one coupled process in a shared geometric reference, rather than as two post-processing steps.

Formally, let $\mathcal{I} = \{I_t\}_{t=1}^{T}$ denote a sequence of images acquired at successive observation times. Each image $I_t$ is processed to generate a binary crack mask $M_t$, which is partitioned into local crack instances using the method described in Section 2.4. Pairwise geometric association then links these local instances and propagates crack identities across consecutive observations. After identity assignment, the corresponding instance skeletons are used to distinguish inherited crack support from newly appeared growth in the shared reference frame.

*2.5.1 Pose-guided crack projection and instance matching*

After geometric projection, the crack masks from the previous and current frames are represented in the same image coordinate system. Geometric projection and segmentation both introduce pixel-level uncertainty. For a one-pixel-wide crack, even a small offset of only a few pixels can eliminate overlap between two observations of the same crack. Therefore, before overlap-based matching, both masks are dilated with the same disk-shaped kernel to create an image-space acceptance band that tolerates minor projection and segmentation errors, as illustrated in Fig. 4. Let $M_{t-1}$ denote the projected instance mask from the previous frame and $M_t$ the candidate instance mask in the current frame. The dilated masks are defined as

$$M'_{t-1} = M_{t-1} \oplus B_{r_{\text{dil}}}, M'_t = M_t \oplus B_{r_{\text{dil}}}, \tag{14}$$

where $\oplus$ denotes morphological dilation and $B_{r_{\text{dil}}}$ is a disk-shaped dilation kernel with radius $r_{dil}$. This operation provides an image-space tolerance for spatial matching and does not represent physical crack widening.

Crack-instance correspondence is determined from the overlap between the dilated masks. For a previous instance $i$ and a current candidate $j$, let $I_{ij}$ denote the intersection area of the two dilated masks, $C_{ij}$ the fraction of the current instance covered by the projected previous instance, and $D_{ij}$ the Dice score between the two dilated masks. A correspondence is accepted when

$$I_{ij} \geq I_{\min}, \quad (C_{ij} \geq C_{\min} \text{ or } D_{ij} \geq D_{\min}) \tag{15}$$

with $I_{\min}$ =1 pixel, $C_{\min}$ =0.02, and $D_{\min}$ =0.05. These low thresholds accommodate the limited overlap commonly observed for thin and fragmented crack masks; they were selected empirically and kept fixed for all scenes (Section 3.4). If multiple previous instances satisfy these conditions for the same current candidate, the previous instance with the largest sum of $C_{ij}$ and $D_{ij}$ is selected, with $I_{ij}$ used as a tie-breaker. This rule provides dilation-tolerant matching for thin and fragmented cracks without introducing an additional learned model or modifying the geometric projection.

*2.5.2 Crack numbering strategy*

Once pairwise crack-instance correspondences have been established, crack identities are propagated chronologically through the image sequence, as illustrated in Fig. 4. In the first frame, each identified crack instance is assigned a unique track identifier. These identifiers provide the initial identity labels for subsequent frames.

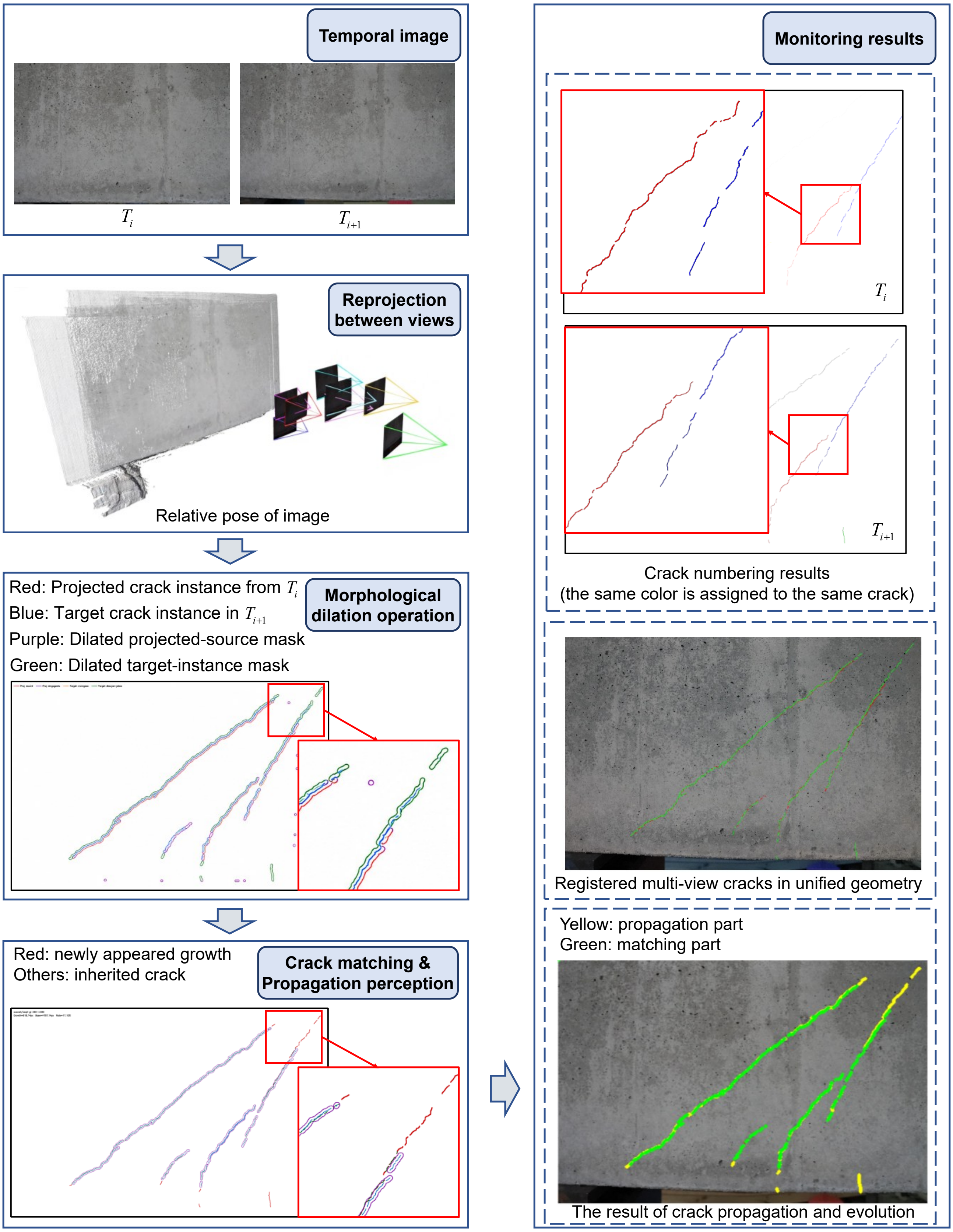

**Fig. 4.** Workflow for crack-instance association, identity propagation, and growth extraction following cross-view geometric alignment.

For each adjacent pair of frames $t$ and $t+1$, the tracked crack instances in frame $t$ are projected into the coordinate system of frame $t+1$. Each crack instance in frame $t+1$ inherits the identifier of the projected previous-frame instance with the highest accepted matching score among the candidates accepted by Eq. (15). Multiple fragments in the current frame may inherit the same identifier when they are associated with the same physical crack. This allows a fragmented or branching crack to retain a consistent identity over time.

Two fallback rules handle weak or missing overlap. First, when no overlap-based association is accepted, an unmatched current instance may inherit the identifier of the nearest projected track within 160 px. Second, in sequences that contain only a single tracked instance, this identity is carried across intervals in which the projection produces no valid pixels, so that short segmentation interruptions do not create spurious new identities. If, after the overlap and fallback rules, a current instance still cannot be associated with any previous track, it is treated as a newly appeared crack and assigned a new track identifier.

After processing the complete scene sequence, predicted instances that have no association with any observation in the scene are assigned to background in the final tracking output. Here, association refers to matching between predicted instances using the rules described above.

Within the annotation convention used in this study, branches associated with the same physical crack root retain one track identifier. Separate identifiers are assigned when the annotations indicate physically distinct cracks. In the visualizations, each track identifier is rendered in a distinct color, and the same color is maintained across frames for the same tracked crack.

### *2.5.3 Pairwise crack-growth extraction and length quantification*

Crack growth is evaluated pairwise between adjacent observations. The skeletons of the same tracked crack instance are first expressed in the coordinate frame of the later observation, as illustrated in Fig. 4. Following this alignment, crack growth is represented by skeleton pixels that newly appear in the later observation. Full-mask comparison would conflate longitudinal crack extension with changes in mask width caused by crack opening, illumination variation, and segmentation noise. The one-pixel-wide skeleton instead provides a centerline representation whose length is additive and comparable across observations.

After identity assignment, let $M_t^k$ denote the mask of tracked crack instance $k$ at observation *t*, and let $S_t^k = \mathrm{Skel}(M_t^k)$ denote its one-pixel-wide skeleton. The previous-stage skeleton registered in the coordinate frame of observation *t*+1 is defined as

$$\tilde{S}_{t\to t+1}^k = T_{t\to t+1}(S_t^k), \tag{16}$$

where $T_{t\to t+1}$ is the pairwise geometric mapping defined in Eq. (1). To distinguish the previous-crack support radius from the new-growth tolerance radius, two separate radii are used. The previous-crack support band is constructed by dilating the registered previous skeleton with the radius $r_p$:

$$B_{p,t}^k = \tilde{S}_{t\to t+1}^k \oplus B_{r_p}. \tag{17}$$

The newly appeared growth of crack instance $k$ between observations *t* and *t*+1 is defined using the additional tolerance $r_g$ as

$$\Delta\hat{S}_{t+1}^k = \{\boldsymbol{x} \in S_{t+1}^k \mid \mathrm{dist}(\boldsymbol{x}, B_{p,t}^k) > r_g\}. \tag{18}$$

Pixels within $r_g$ of the support band are regarded as inherited support.

The extension length of crack instance $k$ between the two consecutive observations is then measured by the length of the extension skeleton

$$\Delta\hat{L}_{k,t} = \mathrm{Len}(\Delta\hat{S}_{t+1}^k). \tag{19}$$

Summing $\Delta\hat{L}_{k,t}$ over all tracked instances gives the predicted crack-length increment over the interval $[t, t+1]$. Thus, crack propagation in this study refers to the newly appeared skeleton length after pairwise registration, allowing structural growth to be distinguished from observation-induced appearance changes.

This design transforms conventional visual matching into a physically interpretable monitoring problem. The fusion of pose-induced geometric registration with morphology-aware correspondence and growth comparison provides a unified framework for cross-view crack numbering and crack-growth tracking.

## 3. Experiment Setup

### 3.1 Data acquisition from staged loading experiments

In in-service concrete structures, crack propagation generally develops over an extended period, while the incremental extension between observations is often very small relative to the existing crack pattern. Capturing a representative crack-growth history may therefore require repeated observations over several months or even years. These practical constraints make it difficult to collect temporally continuous field image sequences of crack propagation for evaluating the proposed method.

The proposed method targets two sources of variation that coexist between successive observations: structural change (crack initiation and growth) and viewpoint change (differences in camera position, orientation, and scale). The tracking algorithm operates on image-level features and does not model the physical mechanisms that drive crack growth. Its key role is to distinguish true structural change from apparent changes caused by viewpoint change. Because the visual signature of a crack in an image is independent of whether the crack formed over minutes under a hydraulic actuator or over years under service loading, any image sequence that simultaneously contains observable crack increments and deliberate viewpoint variation provides a sufficient condition for validating the tracking method. Staged loading satisfies exactly these two requirements: each load increment introduces real, irreversible crack extension, while the camera position and orientation are freely selected at each stage rather than constrained to a fixed station or trajectory, subject only to maintaining overlap around the crack region. Images are acquired at close range, but no fixed camera-to-surface distance is prescribed or measured. Compared to a multi-year field campaign, which would conflate loading history, environmental exposure, surface contamination, and uncontrolled acquisition changes, the laboratory setup isolates the two variables of interest and therefore offers a less confounded validation condition.

The dataset was collected from staged loading tests on three types of reinforced-concrete specimens, deliberately selected so that the dataset covers the cracking scenarios the tracking framework must handle. Pre-notched beams produce a single dominant flexural crack (Mode-I opening), for validating identity-preserving extension tracking of one growing crack. The L-shaped beam induces bifurcating cracks at the re-entrant corner under combined flexural-shear action (mixed Mode I–II), testing branch-aware identity propagation. Deep beams develop multiple diagonal shear cracks (Mode-II dominated), testing parallel multi-crack association and interaction. The specimen geometry, loading positions, and boundary conditions were designed accordingly. Images were acquired using readily available handheld devices, including OnePlus Ace 2 Pro and realme GT5 Pro smartphones and a Sony ILCE-7SM3 camera equipped with an FE 24-70 mm F2.8 GM II lens. The acquisition device and native image resolution were not standardized across the dataset; the native resolutions are reported in Table 1. Accordingly, the evaluation does not assume a single camera model or native image format. As designed, the dataset records Mode-I flexural, mixed-mode branching, and Mode-II shear crack evolution under progressive loading. Each sequence captures the same crack system across multiple stages, including crack initiation, propagation, and interaction, enabling joint analysis of crack identity and newly appeared growth rather than treating them as isolated tasks. Fig. 5 presents a representative example that combines a global specimen view with close-up observations of the same crack region under successive loading stages and changing viewpoints.

### 3.2 Dataset construction

Table 1 summarizes the experimental scenes and their stage-wise observations. The scenes represent three characteristic cracking scenarios: single dominant crack propagation, mixed-mode branching, and multiple-crack interaction. These crack patterns support the evaluation of crack-instance identity tracking and crack-growth tracking under different levels of structural and morphological complexity, while the segmentation front end is held fixed across all scenes.

The dataset contains 53 stage-wise images from eight independent physical specimens; no specimen contributes images to more than one scene. Of these images, 45 adjacent-stage transitions were used for quantitative evaluation. Two levels of annotation were prepared. Pixel-level masks delineate the binary crack regions. Instance labels assign a scene-local identity to each physical crack and preserve that identity across viewpoints and loading stages. Disconnected fragments of the same physical crack may therefore share one annotated identity, whereas different physical cracks receive different identities. All the identities reported in Table 1 are manually annotated crack identities. Fig. 6 presents the original image, binary crack mask, instance-label map, and an enlarged view showing the separation of neighboring crack identities. In the instance-label map, each identity is represented by a distinct color, and repeated use of the same color within a scene denotes the same physical crack. Growth references were constructed from manually annotated crack masks and instance identities using geometric mappings selected through visual assessment of cross-view alignment quality, without restricting the selection to a particular matching method. Where necessary, the input-image set, correspondences, and camera poses were interactively adjusted. The accepted mappings were fixed before evaluation and used consistently across all compared configurations to provide a manually reviewed, geometry-assisted growth reference.

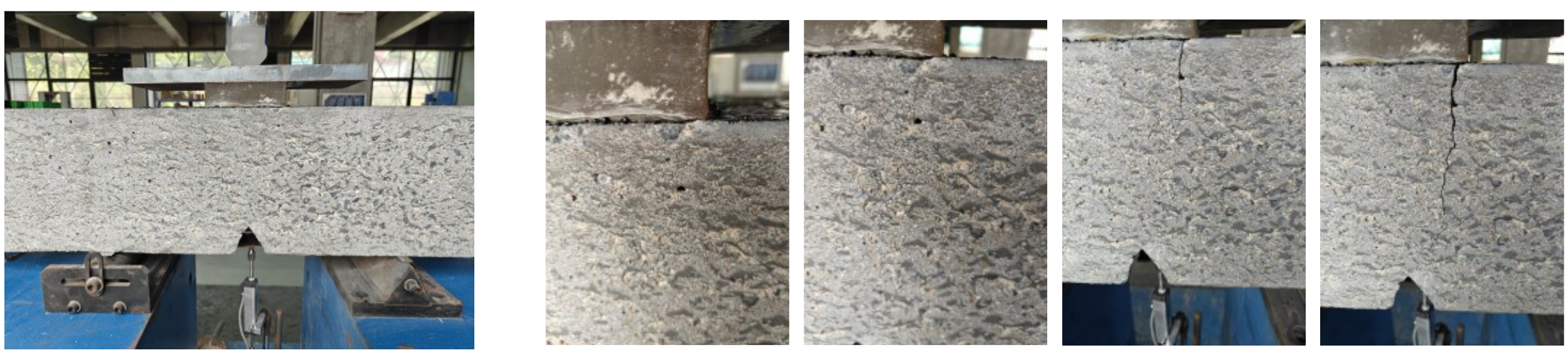

(a) overall view (b) local views of crack sequences

**Fig. 5.** Temporal images of crack propagation in a pre-notched beam

**Table 1.** Summary of the temporal crack-tracking dataset.

| Scene | Specimen and observed crack pattern | Dominant mode | Original image size | Evaluated frames | Number of crack identities |
|---|---|---|---|---|---|
| 1 | Pre-notched beam; single mid-span flexural crack | Mode I | 4096 × 3072 | 4 | 1 |
| 2 | Pre-notched beam; single mid-span flexural crack | Mode I | 4096 × 3072 | 4 | 1 |
| 3 | L-shaped beam; re-entrant-corner bifurcating crack | mixed Mode I–II | 4096 × 2688 | 5 | 1 |
| 4 | L-shaped beam; single span flexural crack | Mode I | 4240 × 2832 | 6 | 1 |
| 5 | L-shaped beam; single span flexural crack | Mode I | 4240 × 2832 | 4 | 1 |
| 6 | Deep beam; multiple edge diagonal shear cracks | Mode II | 4096 × 2688 | 9 | 6 |
| 7 | Deep beam; multiple edge diagonal shear cracks | Mode II | 4240 × 2832 | 11 | 8 |
| 8 | Deep beam; multiple mid-span flexural cracks | Mode I | 4240 × 2832 | 10 | 7 |

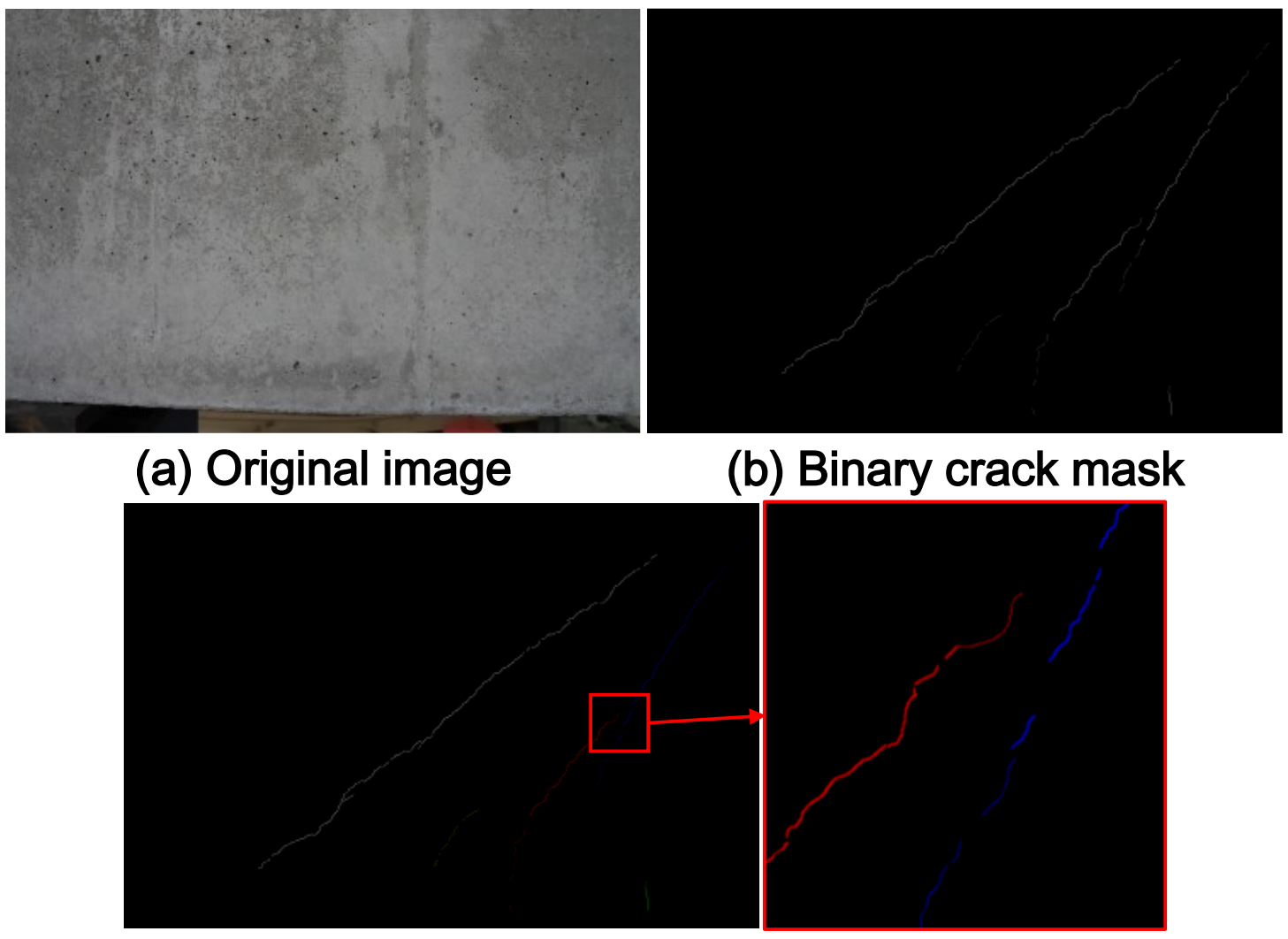

**Fig. 6.** Crack annotation example.

### 3.3 Evaluation metrics

Crack-instance tracking couples instance association with growth quantification. A comprehensive evaluation therefore considers three questions: whether the same physical crack retains a consistent identity across different viewpoints and observation times, whether newly appeared crack growth is localized in the correct spatial position and observation interval, and whether the estimated growth length is consistent with the reference growth labels. Identity correspondence is evaluated using Higher Order Tracking Accuracy (HOTA), which jointly considers detection, association, and localization [33]. HOTA is a robust, physically meaningful, and widely used metric for object tracking. It is also informative for crack monitoring, as temporal identity consistency is a fundamental requirement for tracking individual crack instances. However, HOTA alone is insufficient for evaluating crack-instance tracking. It was originally designed for object-like targets with well-defined spatial extents, such as bounding boxes or segmentation masks of pedestrians and vehicles. Cracks, in contrast, are thin, elongated, irregular, and often topologically complex structures. Their evolution is not limited to simple motion or displacement, but may involve propagation, branching, merging, widening, and partial disappearance. Moreover, HOTA does not determine whether newly observed crack regions correspond to actual crack growth, whether such growth is detected at the correct observation interval, or whether the associated increase in crack length is accurately quantified. Therefore, the proposed evaluation protocol adopts HOTA and the newly developed Crack-Length Change Error (CLCE) as two core metrics to assess identity correspondence and post-association crack-growth quantification, respectively, supplemented by two auxiliary diagnostic metrics. Immediate Growth F1 evaluates growth-length-weighted instance-level growth detection within the same interval. Delayed Growth Recall (DGR)@$K$ measures the fraction of reference growth length detected in its corresponding interval or within the following $K$ observation intervals.

*3.3.1 Growth-length consistency*

To evaluate the tracking and quantification of crack growth, this study proposes the CLCE computed after crack-instance association. The metric compares predicted and verified reference increments in skeleton length for each valid crack instance between adjacent loading stages.

The length of any crack skeleton is measured by a common length operator $\mathrm{Len}(\cdot)$. For a skeleton represented by an ordered set of points $\{\boldsymbol{p}_m\}_{m=1}^{N}$, its discrete length is defined as

$$\mathrm{Len}(P)=\sum_{m=1}^{N-1}\left\|\boldsymbol{p}_{m+1}-\boldsymbol{p}_m\right\|_2 \tag{20}$$

where $\boldsymbol{p}_m$ is the $m$-th skeleton point and $N$ the number of points. For thin and locally discontinuous cracks, $\mathrm{Len}(\cdot)$ can also be approximated by the number of skeleton pixels, which gives a stable estimate for temporal comparison. The length operator in Eq. (20) is used consistently for the extension length in Eq. (19) and for the predicted and reference growth lengths in Eqs. (21) and (22). For an interval $t$, the predicted and verified reference growth lengths of crack instance $k$ are computed from the spatially newly appeared skeleton supports, rather than from the net difference between total skeleton lengths at the two time points:

$$\Delta\hat{L}_{k,t}=\mathrm{Len}(\Delta\hat{S}_{t+1}^{k}) \tag{21}$$

$$\Delta L_{k,t}^{\mathrm{ref}}=\mathrm{Len}(\Delta S_{t+1}^{\mathrm{ref},k}) \tag{22}$$

where $\Delta\hat{S}_{t+1}^{k}$ and $\Delta S_{t+1}^{\mathrm{ref},k}$ denote the predicted and manually verified newly appeared skeleton sets after registration into the same reference view. Because early-stage micro-cracks may have very small reference growth lengths, directly averaging instance-wise relative errors can be dominated by near-zero denominators. The CLCE metric is computed as

$$\mathrm{CLCE}=\frac{\sum_{(k,t)\in\mathcal{V}}\left|\Delta\hat{L}_{k,t}-\Delta L_{k,t}^{\mathrm{ref}}\right|}{\sum_{(k,t)\in\mathcal{V}}\Delta L_{k,t}^{\mathrm{ref}}+\epsilon} \tag{23}$$

where $\mathcal{V}$ is the set of valid crack-instance transitions in the verified reference, and $\epsilon$ is a small positive constant set to $10^{-6}$ to prevent division by zero. In this study, transitions with verified reference growth length $\leq 1$ px are excluded from the aggregation to prevent near-zero increments from dominating the aggregate ratio. Thus, CLCE measures instance-level crack-length change error after crack-instance association. It does not assess the spatial localization of predicted growth.

*3.3.2 Growth localization and delayed crack-growth detection*

Instance-level growth detection should penalize both missed growth and spurious growth, which corresponds precisely to growth-length-weighted recall and precision. An F1 score therefore summarizes instance-level growth detection performance in one number, evaluated length-wise so that larger growth increments contribute proportionally to their monitoring importance. Immediate Growth F1 is reported as an auxiliary same-interval instance-level growth-detection diagnostic, comparing predicted and reference growth skeletons in the same pair interval using the growth support band.

Let $P_t$ and $G_t$ denote the predicted and reference growth skeletons at interval *t*. For each crack instance, its growth skeleton comprises the newly appeared skeleton pixels within one interval, including disconnected portions; growth skeletons with growth length at most 1 px are excluded. Let $L_t^P$ and $L_t^G$ denote the total predicted and reference growth lengths, and $M_t^P$ and $M_t^G$ the corresponding totals over matched growth skeletons. The pairwise support-band F1 is the harmonic mean of the fractions of each skeleton's pixels covered by the other's support band, and is zero if either skeleton is empty or both fractions are zero. A growth skeleton is matched if at least one counterpart in the same scene and interval has a bidirectional skeleton support-band F1 of at least 0.10, using a 16 px radius. No one-to-one constraint is imposed, and each matched growth skeleton contributes its full length once on its own side. Precision, recall, and F1 are defined as

$$\mathrm{Pre}_0 = \frac{\sum_t M_t^P}{\sum_t L_t^P + \epsilon}, \tag{24}$$

$$\mathrm{Rec}_0 = \frac{\sum_t M_t^G}{\sum_t L_t^G + \epsilon} \tag{25}$$

$$\mathrm{F1}_{\mathrm{IG}} = \frac{2\mathrm{Pre}_0\mathrm{Rec}_0}{\mathrm{Pre}_0 + \mathrm{Rec}_0 + \epsilon}. \tag{26}$$

where $\mathrm{F1}_{\mathrm{IG}}$ denotes the Immediate Growth F1 score, and $\epsilon$ prevents division by zero. Since $M_t^P \le L_t^P$ and $M_t^G \le L_t^G$, both precision and recall, and therefore $\mathrm{F1}_{\mathrm{IG}}$, are bounded between zero and one.

To assess segmentation-related delayed detection without treating temporal tolerance as the main accuracy criterion, one additional quantity is reported. Early micro-cracks may be detected only after they widen, so reference growth missed in its corresponding interval may be detected in a subsequent observation. The parameter *K* defines the number of subsequent pair intervals considered. DGR@*K* measures the fraction of reference growth length detected either in its corresponding interval or within the next *K* pair intervals. For each reference growth segment $G_{k,t} \subseteq G_t$, candidate predictions in the corresponding interval and the subsequent *K* pair intervals are examined in increasing order of delay, with the search truncated at the end of the sequence. Before comparison, the reference segment is projected into the target frame of the candidate prediction using the consecutive pairwise mappings. At each delay, one-to-one assignment is performed using the bidirectional support-band F1, with a minimum pairwise matching score of 0.10. Once assigned, a prediction cannot be reused for another reference growth segment:

$$\mathrm{DGR}@K = \frac{\sum_{k,t} w_{k,t} c_{k,t}^{(K)}}{\sum_{k,t} w_{k,t} + \epsilon}, \tag{27}$$

where *k* denotes the crack-instance index, *t* denotes the pair interval, $w_{k,t}$ is the original measured length of the reference growth skeleton, and $c_{k,t}^{(K)}$ is the reference-side coverage obtained from the earliest one-to-one match within the allowable *K*-interval window. Specifically, $c_{k,t}^{(K)}$ is the fraction of pixels of the projected reference growth skeleton lying within the support band of its assigned prediction. Each prediction is used at most once, and an unmatched reference segment contributes zero. Since $0 \le c_{k,t}^{(K)} \le 1$, DGR@*K* is also bounded between zero and one.

In this study, Immediate Growth F1 and DGR@*K* are reported as auxiliary diagnostics: the former

evaluates same-interval instance-level growth localization, whereas the latter evaluates reference-growth coverage in the original interval or within the following *K* observation intervals.

*3.3.3 Crack-instance identity correspondence*

Crack-instance identity preservation was evaluated using the standard HOTA formulation [33] on the final tracking outputs defined in Section 2.5.2. All ground-truth instances were retained, and unmatched final predictions and ground-truth instances were counted as false positives and false negatives, respectively. For an intersection-over-union (*IoU*) threshold $\tau$, the overall HOTA score is computed as

$$\mathrm{HOTA}(\tau) = \sqrt{\mathrm{DetA}(\tau) \times \mathrm{AssA}(\tau)}, \tag{28}$$

where $\mathrm{DetA}(\tau)$ and $\mathrm{AssA}(\tau)$ denote detection and association accuracies, respectively. The detection term is defined as

$$\mathrm{DetA}(\tau) = \frac{|\mathrm{TP}(\tau)|}{|\mathrm{TP}(\tau)| + |\mathrm{FN}(\tau)| + |\mathrm{FP}(\tau)|}, \tag{29}$$

where $\mathrm{TP}(\tau)$, $\mathrm{FN}(\tau)$, and $\mathrm{FP}(\tau)$ denote true-positive, false-negative, and false-positive crack-instance matches, respectively. The association accuracy, $\mathrm{AssA}(\tau)$, quantifies the consistency of crack-instance identities across viewpoints and observation times and is defined as

$$\mathrm{AssA}(\tau) = \frac{1}{|\mathrm{TP}(\tau)|} \sum_{c \in \mathrm{TP}(\tau)} \frac{|\mathrm{TPA}(c)|}{|\mathrm{TPA}(c)| + |\mathrm{FNA}(c)| + |\mathrm{FPA}(c)|}, \tag{30}$$

where $\mathrm{TPA}(c)$, $\mathrm{FNA}(c)$, and $\mathrm{FPA}(c)$ are the true-positive, false-negative, and false-positive associations evaluated for a matched crack instance $c$, and $\mathrm{LocA}(\tau)$ is the average localization overlap of matched instances, reported together with the HOTA curves. HOTA was computed on the final tracking outputs defined in Section 2.5.2 using 19 *IoU* thresholds ranging from 0.05 to 0.95 [33]. Accordingly, HOTA(0) denotes the score at the first threshold ( $\tau = 0.05$ ), rather than a threshold defined specifically for this study. It serves as a loose identity-correspondence measure, whereas mean

HOTA provides a stricter assessment of overlap stability. For thin crack instances, a small lateral displacement caused by segmentation uncertainty or projection drift can substantially reduce mask overlap even when the two observations correspond to the same physical crack. Therefore, the *IoU* threshold $\tau$ should be interpreted as a strictness control on projected instance overlap in the reference view, rather than as a crack width, a physical damage threshold, or a direct geometric-error tolerance.

Taken together, HOTA evaluates crack-instance identity correspondence, CLCE evaluates growth-length consistency after instance association, Immediate Growth F1 evaluates same-interval instance-level growth localization, and DGR@*K* diagnoses delayed detection.

### 3.4 Implementation details

The implementation follows the four modules introduced in Section 2.1. The principal parameters are summarized in Table 2 and were kept fixed across all eight scenes.

For image crack segmentation, crack masks were generated using the DeepLabv3+ pipeline reported in [30]. The checkpoint was trained and selected in that study; no image from the eight temporal scenes evaluated here was used for its training or fine-tuning. It was applied without further training and remained fixed throughout the tracking experiments. The training, validation, and test data used in [30] do not share any physical specimens with the eight temporal scenes evaluated here. These temporal scenes were used only for tracking and growth evaluation. During inference, full-resolution images were divided into non-overlapping 448 × 448 patches. The predictions were stitched and aligned with the cropped coordinate system used for tracking. The same predicted masks were used for all tracking and growth analyses.

For geometric structure estimation, a pretrained MASt3R model [27, 28] was applied to each adjacent observation pair. Its 3D point maps, valid-pixel masks, camera poses, and focal lengths were used to construct the source-to-target mapping defined in Eq. (1). Image dimensions and cropping offsets were checked before projection to ensure consistency between the masks and geometric data.

For crack instance identification, each predicted mask is partitioned into 8-connected components and skeletonized. Primitive branches are extracted at endpoints and junctions. The separation criterion uses dominant directions estimated from the complete skeleton extent of each branch. When both direction estimates satisfy the reliability conditions in Table 2, branch pairs with an acute angle of at least 45° are assigned a hard-separation constraint, as defined in Eqs. (12)–(13). Junction links are processed first, after which remaining fragment-pair candidates are evaluated using the core merge conditions in Eq. (11) and the additional endpoint-axis criteria described in Section 2.4. Constrained union-find rejects any union that would reconnect a hard-separated branch pair. Foreground pixels are assigned to primitive-branch skeletons by nearest-seed label propagation and relabeled according to the final groups. No component-area filtering is applied, and all predicted foreground pixels are retained at this stage. During crack-instance association, source instances were projected into the target observation and matched according to Eq. (15). The highest-scoring valid candidate was selected, while multiple target fragments could inherit the same identity. The fallback rules described in Section 2.5.2 were used to bridge short segmentation interruptions. The maximum distance for nearest-track fallback association was set to 160 px in the processed image coordinates. Distance was measured as the minimum separation between the foreground pixels of the unmatched instance and those of the projected track. The final instance-label maps were generated using the temporal association and candidate-acceptance rules described in Section 2.5.2. The original binary segmentation masks were retained unchanged as intermediate outputs.

Crack growth was extracted after pairwise alignment according to Eqs. (16)–(19). The projected previous-crack skeleton was dilated using the support radius ($r_p$). Later-stage skeleton pixels farther than the tolerance ($r_g$) from this support band were classified as newly appeared growth. Both radii were set to 16 px and represent image-space tolerances for projection error and mask discontinuity rather than physical crack width.

The experiments were conducted on Ubuntu 22.04 LTS using an Intel i9-13900K CPU, 128 GB

of RAM, and an NVIDIA RTX 4090 GPU with 24 GB of memory. The learning-based components were implemented in PyTorch 2.0.0 with CUDA 11.5.119.

## 4. Experimental Results and Discussion

### 4.1 Qualitative analysis of crack-instance tracking

As illustrated in Figs. 7–9, the qualitative examples examine whether the same physical crack retains its identity when both viewpoint and loading stage change, and whether inherited crack support can be separated from newly appeared growth. In the visualizations, crack segments inherited from the previous stage are shown in green, whereas newly observed growth or branch segments are highlighted in yellow.

**Table 2.** Implementation parameters used in the primary experiments.

| Module | Parameter | Symbol | Value |
|---|---|---|---|
| Instance identification | Direction-reliability limits | — | 4 pixels; 8 px principal-axis extent |
| Instance identification | Maximum candidate endpoint gap | $g_{\mathrm{merge}}$ | 260 px |
| Instance identification | Close-gap breakpoint | $g_{\mathrm{close}}$ | 28 px |
| Instance identification | Directional-tolerance parameters | $\theta_0$, $\Delta\theta_{\max}$, $\alpha$ | 58°; 23°; 0.12°/px |
| Instance identification | Base normal-offset tolerance and distance coefficient | $d_0$, $\lambda_n$ | 50 px; 0.04 |
| Instance identification | Minimum endpoint-axis alignment for moderate and large gaps | — | 0.18; 0.34 |
| Instance identification | Gap limits for unreliable directions | — | 24, 65, and 143 px |
| Instance identification | Branch-direction separation threshold | $\theta_{\mathrm{split}}$ | 45° |
| Instance association | Projected-source and target-instance dilation radius | $r_{\mathrm{dil}}$ | 16 px |
| Instance association | Minimum intersection | $I_{\min}$ | 1 pixel |
| Instance association | Minimum target-instance coverage | $C_{\min}$ | 0.02 |
| Instance association | Minimum Dice score | $D_{\min}$ | 0.05 |
| Growth extraction | Previous-crack support radius | $r_p$ | 16 px |
| Growth extraction | New-growth tolerance radius | $r_g$ | 16 px |

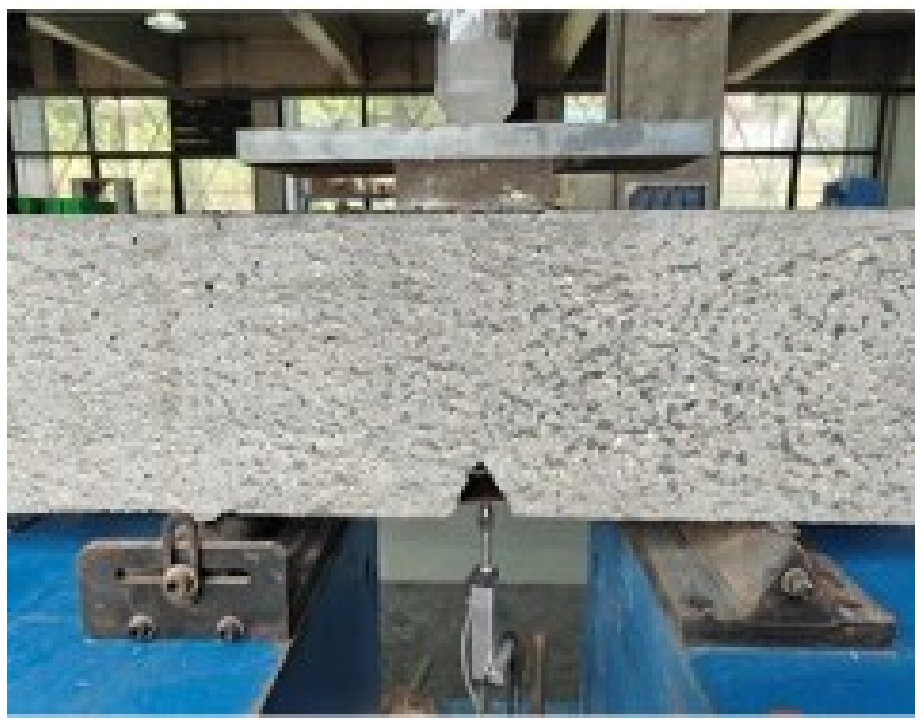

(a) A pre-notched beam

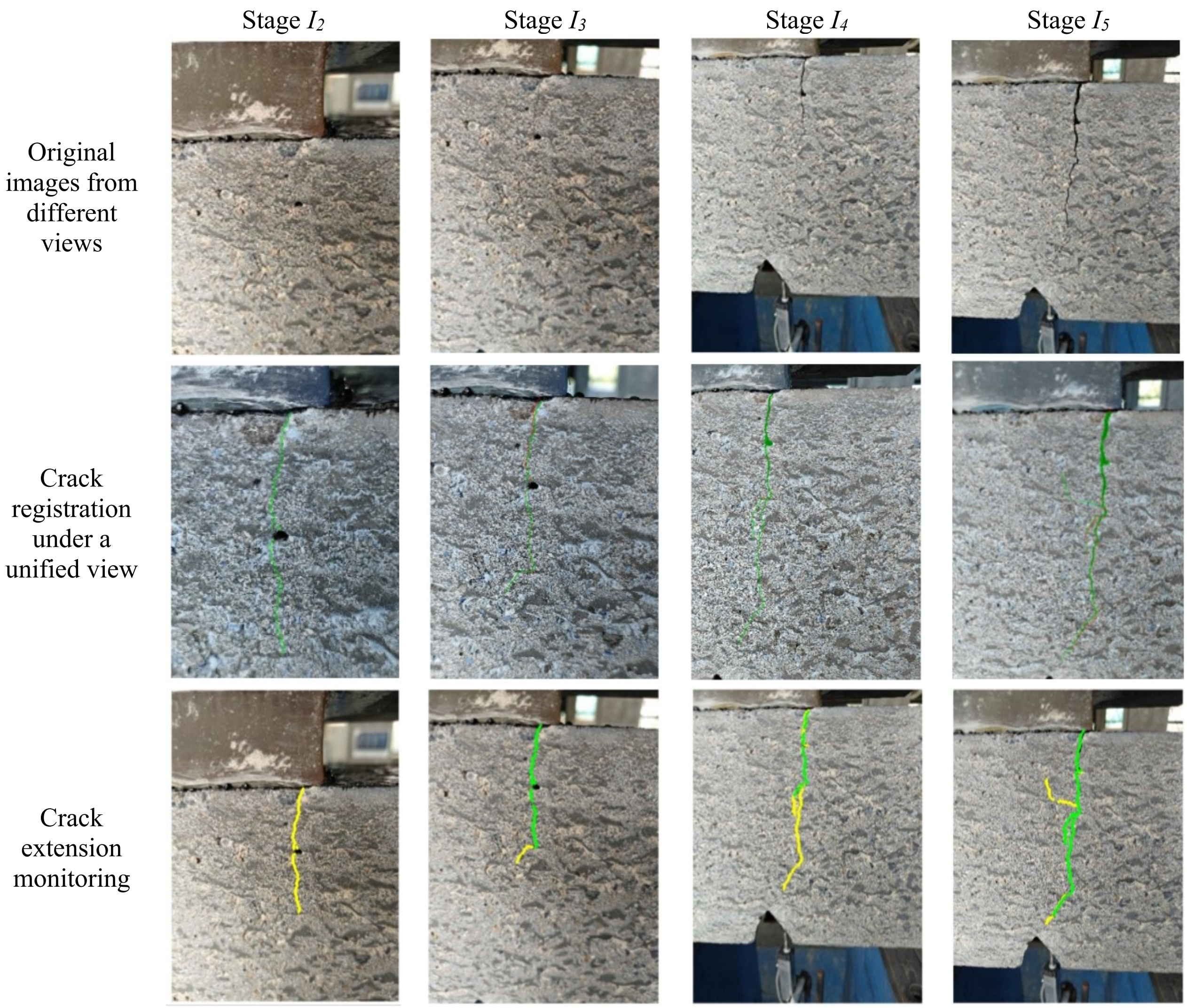

(b) Crack growth tracking

**Fig. 7.** Qualitative results of cross-view crack registration and growth estimation for a pre-notched beam across successive loading stages.

The qualitative results are organized by crack-evolution scenario. For the single-dominant-crack scenario, Fig. 7 presents the result for a pre-notched beam. After registration in the unified reference view, green denotes crack segments inherited from the previous stage, whereas yellow denotes newly observed extension segments. The result shows that the main flexural crack gradually lengthens along a stable path despite changes in camera viewpoint. The framework preserves its identity across stages while separating newly extended segments from the inherited crack trunk.

For a branching crack, the L-shaped beam result in Fig. 8 shows a short local crack developing into a turning and branching pattern near the re-entrant corner. The numbering result keeps the parent crack identity while marking newly formed branches explicitly. For a multi-crack shear scenario, the deep-beam result in Fig. 9 shows multiple shear cracks emerging, propagating in parallel, and interacting locally as loading increases. The unified-view mapping distinguishes pre-existing cracks from newly appeared segments while preserving the major crack identities across stages.

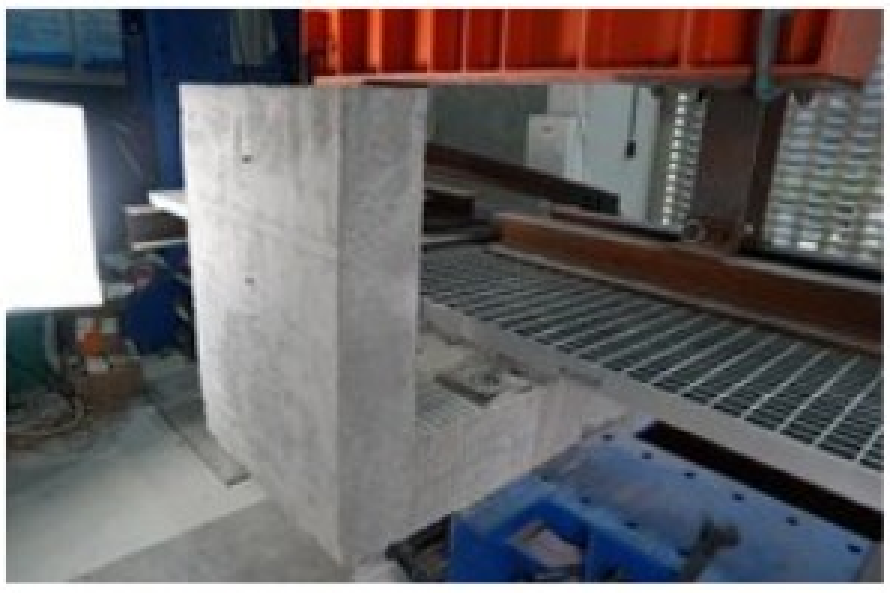

(a) An L-shaped beam

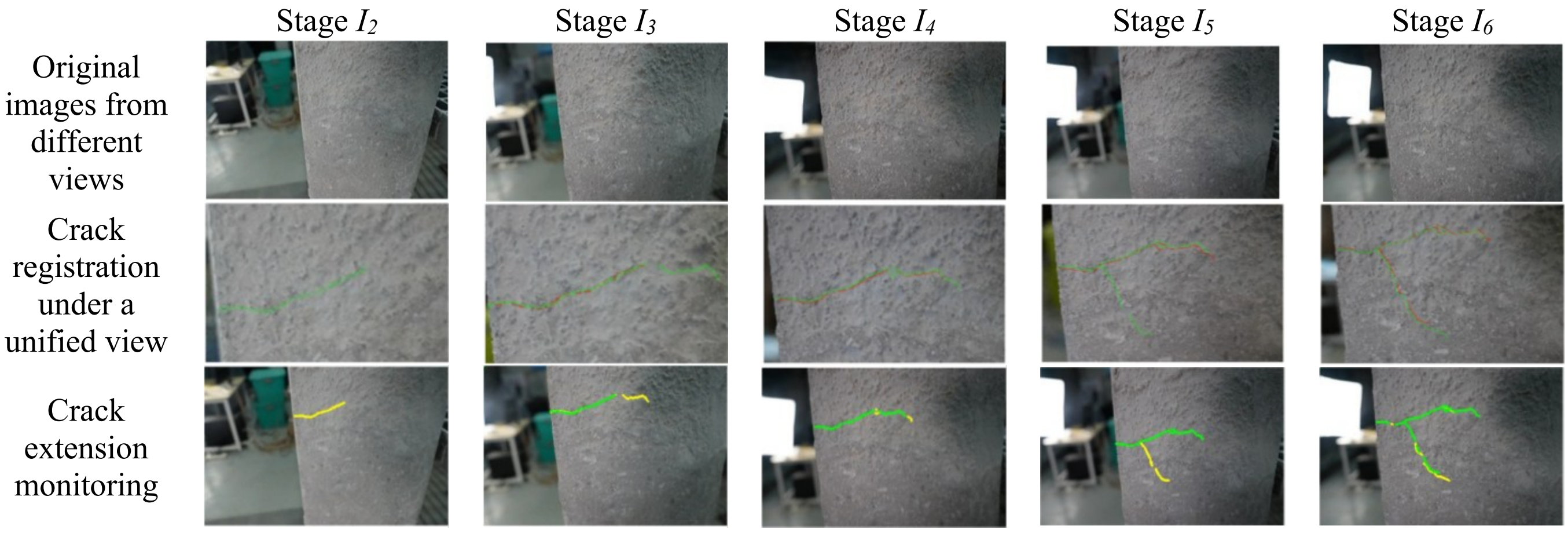

(b) Crack growth tracking

**Fig. 8.** Qualitative results of cross-view crack registration and growth estimation for an L-shaped beam across successive loading stages.

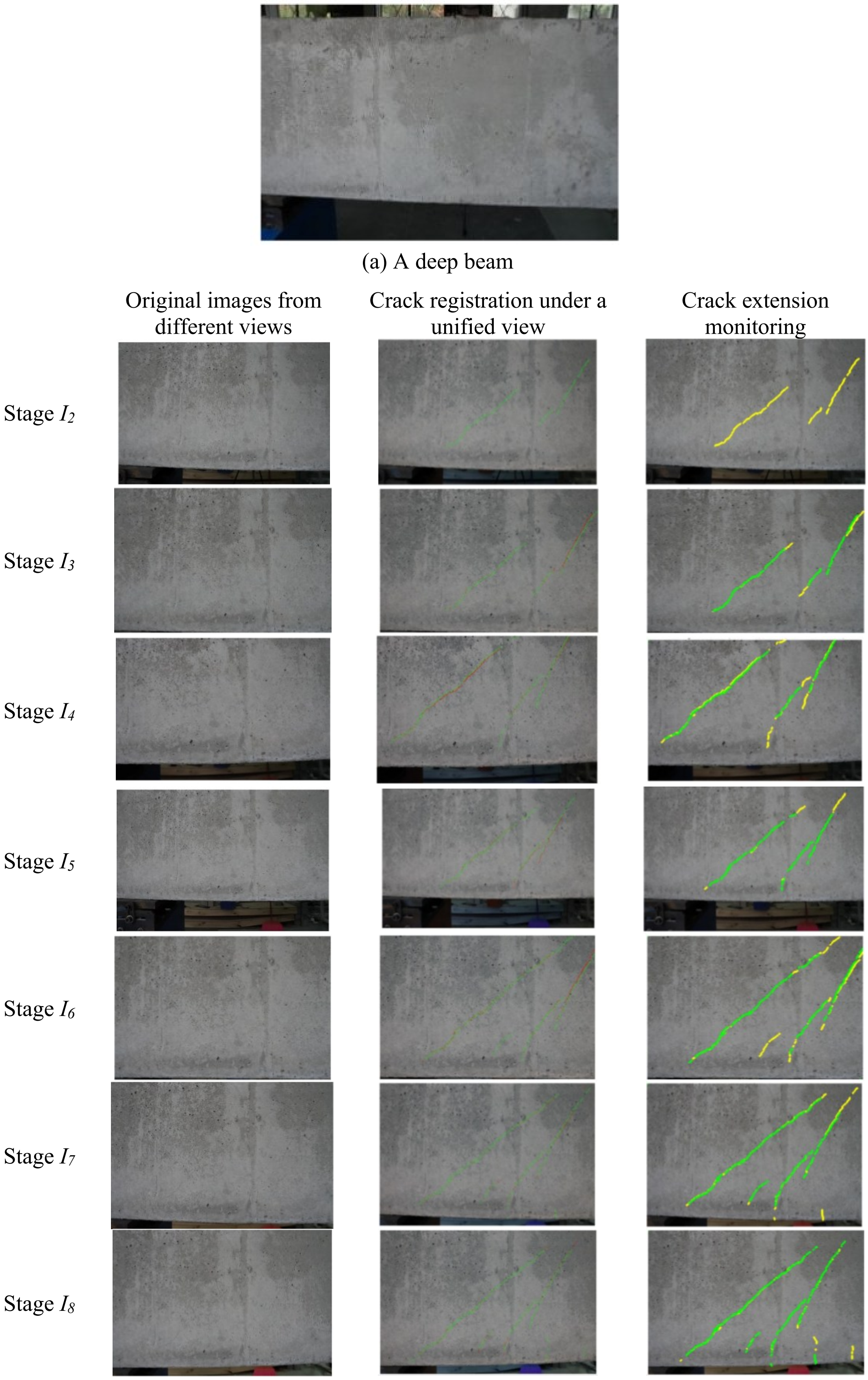

(a) A deep beam

(b) Crack growth tracking

**Fig. 9.** Qualitative results of cross-view crack registration and growth estimation for a deep beam across successive loading stages.

## 4.2 HOTA-based identity-correspondence analysis

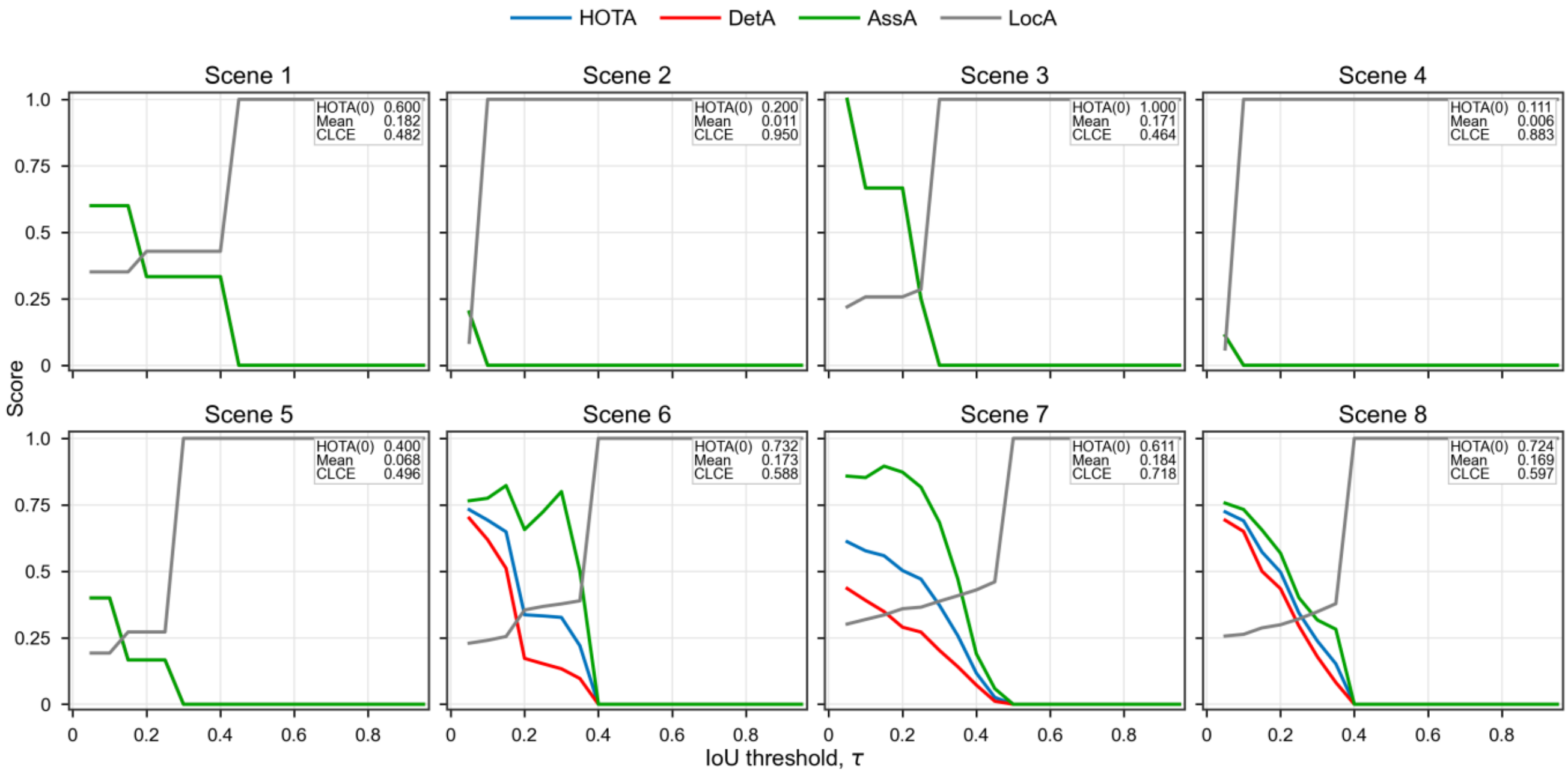

**Fig. 10.** HOTA, DetA, AssA, and LocA as functions of the IoU threshold $\tau$ for the eight experimental scenes.

**Table 3.** Scene-level crack-growth perception and tracking accuracy of the final pipeline.

| Metric | Scene 1 | Scene 2 | Scene 3 | Scene 4 | Scene 5 | Scene 6 | Scene 7 | Scene 8 |
|---|---|---|---|---|---|---|---|---|
| HOTA(0) | 0.600 | 0.200 | 1.000 | 0.111 | 0.400 | 0.732 | 0.611 | 0.724 |
| Mean HOTA | 0.182 | 0.011 | 0.171 | 0.006 | 0.068 | 0.173 | 0.184 | 0.169 |
| CLCE | 0.482 | 0.950 | 0.464 | 0.883 | 0.496 | 0.588 | 0.718 | 0.597 |

The HOTA, DetA, and AssA terms used in this analysis are defined in Eqs. (28)–(30). To assess crack-instance identity preservation after projection into the unified reference view, Fig. 10 presents HOTA, DetA, AssA, and LocA across the $IoU$ threshold $\tau$ for the eight experimental scenes, while the first two rows of Table 3 summarize the corresponding HOTA(0) and mean HOTA values.

In this context, $\tau$ specifies the required $IoU$ between projected crack instances in the unified reference view. As defined in Section 3.3.3, HOTA(0) denotes the score at the first evaluated threshold, $\tau = 0.05$, whereas mean HOTA averages the HOTA scores over all 19 thresholds from 0.05 to 0.95. As $\tau$ increases, the metric progressively demands stricter geometric agreement, which becomes especially challenging for pixel-wide cracks. Under this loose criterion, HOTA(0) ranges from 0.111 to 1.000 across Scenes 1–8, indicating that loose identity correspondence varies substantially among the scenes.

This gap should not be attributed to a single source. It reflects the coupled influence of front-end crack segmentation quality and geometric projection stability. Because the temporal images were collected from the moment cracks first appeared, early-stage cracks are often only about one pixel wide, with weak grayscale contrast, blurred boundaries, and poor continuity. Incomplete masks at this stage lead to fragmented instances and reduced projected overlap in the unified reference view. Meanwhile, texture-poor regions, large viewpoint changes, and partial occlusion may further amplify projection drift, causing substantial reductions in $\mathrm{DetA}(\tau)$ and $\mathrm{AssA}(\tau)$ under strict *IoU* thresholds even when the physical crack identity is preserved. This overlap sensitivity limits the use of HOTA as a standalone metric for thin, deformable, and growing crack instances.

This interpretation is also supported by the HOTA curves in Fig. 10. In most scenes, HOTA, DetA, and AssA remain non-zero only within a relatively small threshold range and then rapidly decay toward zero as $\tau$ increases. The relatively high HOTA(0) values indicate that the proposed framework can establish loose crack-instance correspondence across viewpoints and successive stages of crack development under the loose overlap criterion, whereas the lower mean HOTA values show that strict spatial overlap remains challenging for early-stage and weakly visible cracks. This is why HOTA is used here as an identity-correspondence diagnostic rather than as a complete measure of crack-growth tracking. Immediate Growth F1 evaluates same-interval instance-level growth localization, DGR@*K* quantifies delayed detection, and CLCE measures growth-length consistency after instance association.

### 4.3 Crack-growth tracking analysis

Effective crack-growth tracking requires not only detecting newly emerged growth segments but also localizing them in the correct temporal interval and measuring their length consistently. These capabilities are evaluated from two complementary perspectives: growth-length estimation accuracy and temporal detection delay. Specifically, the last row of Table 3 reports the CLCE as the growth-length error of the final pipeline, while Fig. 11 diagnoses the temporal-delay behavior of growth detection.

**Growth-Length Estimation Accuracy.** The pipeline achieves a pooled instance-level CLCE of 0.753 over 115 valid reference crack-instance transitions across all evaluated image transitions. The scene-level results are reported in Table 3, with individual scene values ranging from 0.464 to 0.950. The notable variance across scenes reveals distinct failure modes. Scenes with lower CLCE (e.g., ~0.5) benefit from stable instance-level correspondence and clean mask boundaries, allowing the inherited crack support to faithfully reflect true geometric changes. In contrast, higher CLCE values (>0.9) primarily stem from three factors, each with a clear mechanistic origin. (1) Severe mask fragmentation disrupts continuous length measurement, because the length operator $\mathrm{Len}(\bullet)$ sums segment lengths along an ordered skeleton path: fragmentation breaks this path into disconnected pieces, so the gaps between fragments are no longer counted faithfully and the measured length deviates from the physical crack trajectory. (2) Local projection drift between adjacent stages misaligns the inherited crack support, because a pixel-level offset shifts the registered previous skeleton relative to the current one; within the support-band comparison, this offset is then misclassified as either newly appeared growth or missing inherited support, biasing the measured length difference in either direction. (3) Insufficient valid reference transitions amplify statistical fluctuations, because CLCE is a length-weighted ratio aggregated over the valid transition set: once near-zero-growth transitions are excluded, only a few transitions may remain, and a single transition with a large error can then dominate the ratio and destabilize the scene-level value. These results indicate that while the pipeline captures the macroscopic trend of crack elongation, fine-grained length consistency remains sensitive to segmentation quality and registration stability.

**Temporal Delayed-Detection Analysis.** Immediate Growth F1 evaluates whether newly appeared growth is detected within the corresponding adjacent-stage transition. Across all transitions, the full method obtains an Immediate Growth Precision of 0.759, Recall of 0.624, and F1 of 0.685. To examine whether growth missed at its first appearance is recovered shortly afterward, Fig. 11 reports DGR@*K*. For this diagnostic, each reference growth skeleton may be matched in its original interval or within the following *K* pair intervals using the earliest one-to-one matching rule defined in Section 3.3.2. The resulting DGR@*K* values are 0.303, 0.313, 0.317, and 0.319 for *K* =0,1,2, and 3, respectively. Immediate Growth Recall counts the full length of each matched reference growth skeleton, whereas DGR@0 counts only its covered fraction under one-to-one matching.

This delayed-recovery analysis complements CLCE, which measures growth-magnitude consistency but does not determine whether growth is localized in the correct interval. Early hairline cracks may be missed at their first appearance and detected only after widening. However, the increase from 0.303 at *K*=0 to 0.317 at *K*=2 corresponds to only 1.4 percentage points of additional recovered reference growth. Thus, allowing two additional intervals provides limited extra reference-growth coverage under the adopted assignment rule and available observations.

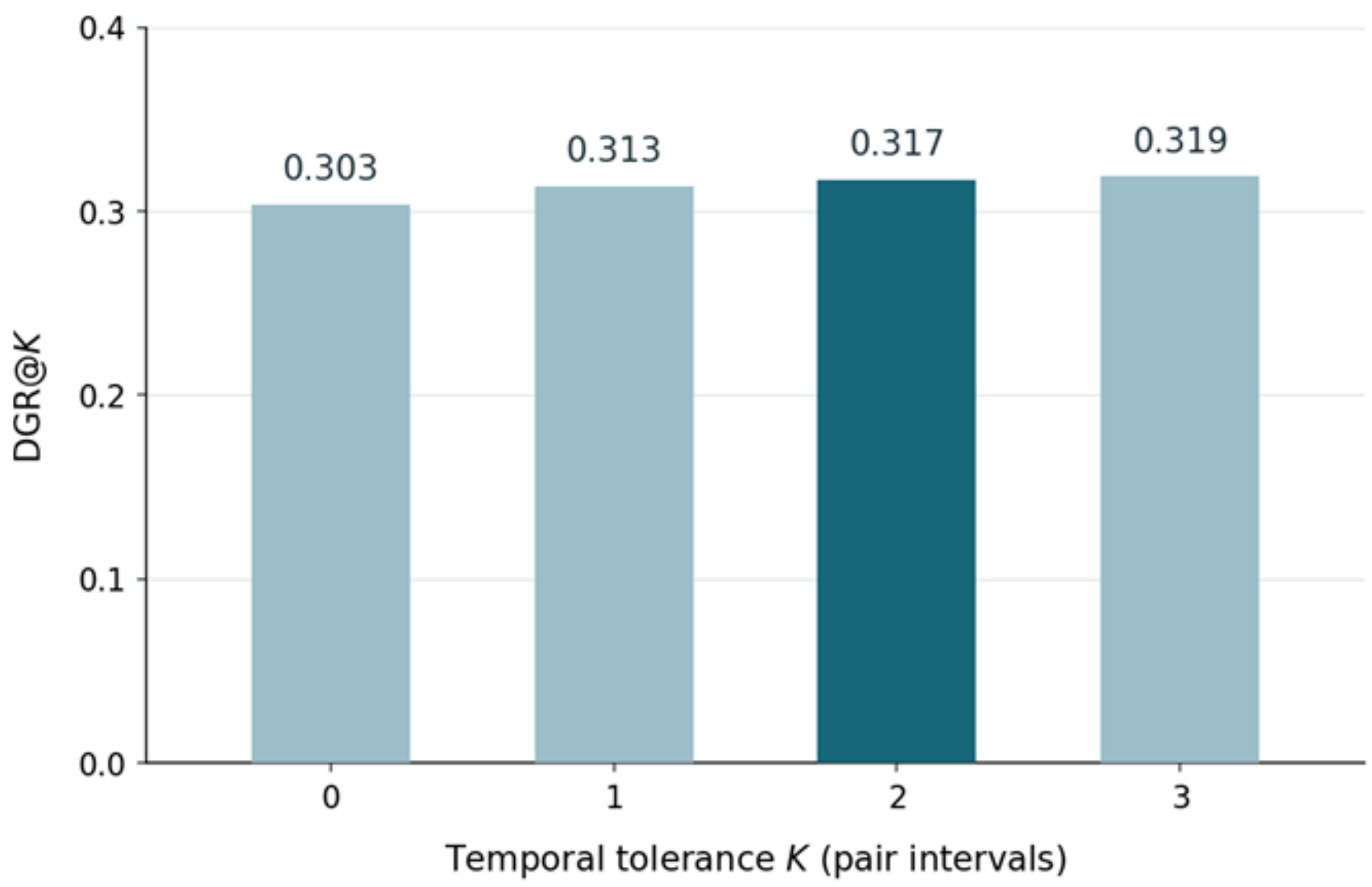

**Fig. 11.** Delayed Growth Recall for different values of *K*.

**Coupling Spatial and Temporal Performance.** Length estimation and temporal detection are inherently coupled. A scene may achieve acceptable crack correspondence under a relaxed overlap criterion yet exhibit high same-step CLCE if minor geometric distortions accumulate between stages. Conversely, a low CLCE does not guarantee correct spatial localization if the predicted growth length is accurate but spatially shifted. Ultimately, robust crack-growth tracking relies on a stable chain of crack identity maintenance, precise projection geometry, and the system's ability to resolve early-stage micro-cracks, each of which presents a distinct avenue for future improvement.

## 4.4 Ablation study

The main ablations examine the effects of geometric registration and explicit crack-instance representation on tracking and growth quantification. To ensure consistent scoring across configurations, all growth metrics in this subsection are computed against the same fixed, manually reviewed common growth reference. The projection results used to construct the reference were manually checked for geometric consistency before being accepted. The resulting reference was fixed before evaluation and applied identically to all compared methods. Note that near-zero-growth transitions (reference growth $\leq 1$ px) are excluded from the CLCE aggregation to avoid statistical instability. The whole-mask variant treats all foreground pixels of each nonempty input mask as one instance, while retaining the full method’s binary masks, geometry, and tracking settings.

### *4.4.1 Module-Level Ablation*

Table 4 summarizes the ablation results over all 45 evaluated transitions. Replacing the pairwise-local MASt3R geometry used in the proposed pipeline with an ASpanFormer homography leaves HOTA(0) essentially unchanged (0.547 versus 0.548), while CLCE increases from 0.753 to 0.769. The near-equivalent HOTA(0) values indicate that the tracking logic is not rigidly tied to one geometry backend, although MASt3R yields a slightly lower growth-length error under the adopted common reference.

**Table 4.** Module-level ablation results.

| Method | Main change | HOTA(0) ↑ | Mean AssA ↑ | Immediate Growth F1 ↑ | CLCE ↓ |
|---|---|---|---|---|---|
| Full method | MASt3R + full crack instantiation | 0.547 | 0.150 | 0.685 | 0.753 |
| ASpanFormer registration | Replace dense geometry with 2D homography | 0.548 | 0.150 | 0.663 | 0.769 |
| Scene-level geometry | Use scene-level rather than adjacent-pair geometry | 0.484 | 0.131 | 0.403 | 1.844 |
| Whole-mask instance | Treat all foreground as one instance | 0.469 | 0.119 | 0.747 | 0.973 |

The importance of adjacent-pair geometry becomes clearer when scene-level geometry is used instead. Using scene-level geometry alone reduces HOTA(0) from 0.547 to 0.484 and Immediate Growth F1 from 0.685 to 0.403, while increasing CLCE from 0.753 to 1.844. These changes show that pairwise-local geometry contributes to both crack-identity correspondence and growth-length estimation.

The whole-mask baseline evaluates the role of explicit crack-instance representation. Compared with the full method, the whole-mask baseline reduces HOTA(0) from 0.547 to 0.469 and Mean AssA from 0.150 to 0.119, and increases CLCE from 0.753 to 0.973. Immediate Growth F1 rises from 0.685 to 0.747, but this instance-level diagnostic does not require correct crack identity and therefore does not establish more accurate crack-wise growth attribution. These results support explicit instance representation for crack-wise tracking and growth attribution under the evaluated protocol.

*4.4.2 Geometry Matcher Comparison*

To ensure consistent scoring, Table 5 compares MASt3R and ASpanFormer on all 45 evaluated transitions. Valid mapping coverage denotes the fraction of target-canvas pixels with finite mapped coordinates located within the source-image bounds and is reported as a comparability check rather than a measure of projection accuracy. Under essentially matched valid mapping coverage (0.774 for ASpanFormer and 0.765 for MASt3R), MASt3R achieves a higher Immediate Growth F1 (0.685 versus 0.663) and a lower CLCE (0.753 versus 0.769). These results indicate that, under the adopted common reference, the MASt3R-based geometry front end provides improved downstream growth-tracking performance, with better same-interval instance-level growth localization under the adopted matching criterion and lower growth-length error. Fig. 12 presents representative coordinate-audited overlays from these geometry configurations.

*4.4.3 Sensitivity of Growth Support Band*

The growth support band acts as an image-space tolerance to absorb projection drift and segmentation discontinuities, rather than representing physical crack width. Fig. 13 analyzes the sensitivity of CLCE to the previous-crack support radius ($r_p$) and the new-growth tolerance radius ($r_g$). The red box marks the default 16 px setting for both radii. The heatmap shows that CLCE increases when both radii are too small, as the resulting spatial tolerance is insufficient to accommodate projection-induced misalignment between corresponding crack segments. Conversely, the system remains stable within the range of 14–16 px. This stability confirms that the selected default radius settings provide a robust buffer for thin-crack projection without introducing excessive false positives.

**Table 5**. Geometry-front-end comparison over all transitions.

| Matcher | Transitions | Valid instance transitions | Immediate Growth F1 | CLCE | Valid mapping coverage |
|---|---|---|---|---|---|
| ASpanFormer homography | 45 | 115 | 0.663 | 0.769 | 0.774 |
| MASt3R pairwise-local | 45 | 115 | 0.685 | 0.753 | 0.765 |

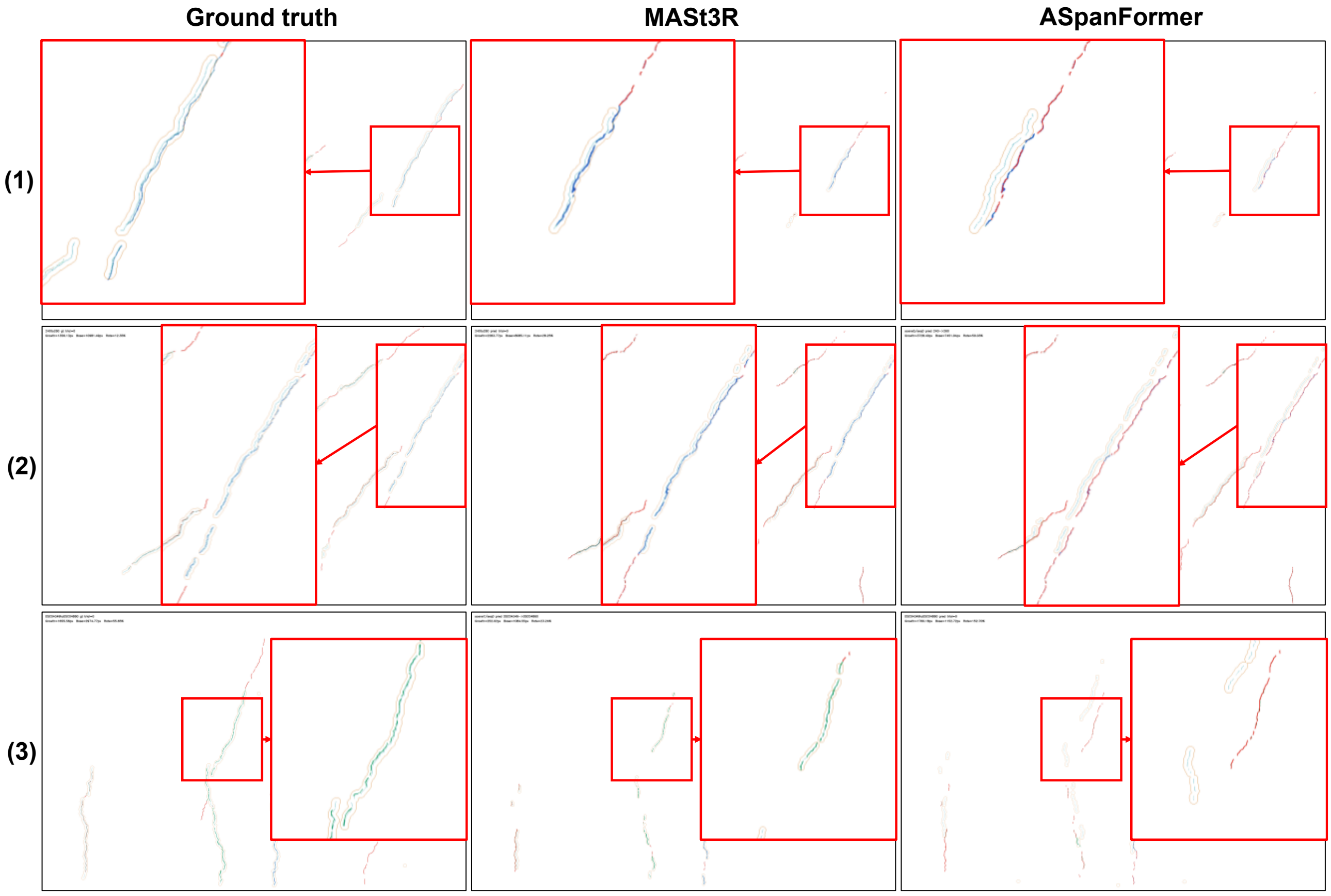

**Fig. 12.** Qualitative comparison of reference growth with MASt3R and ASpanFormer projections.

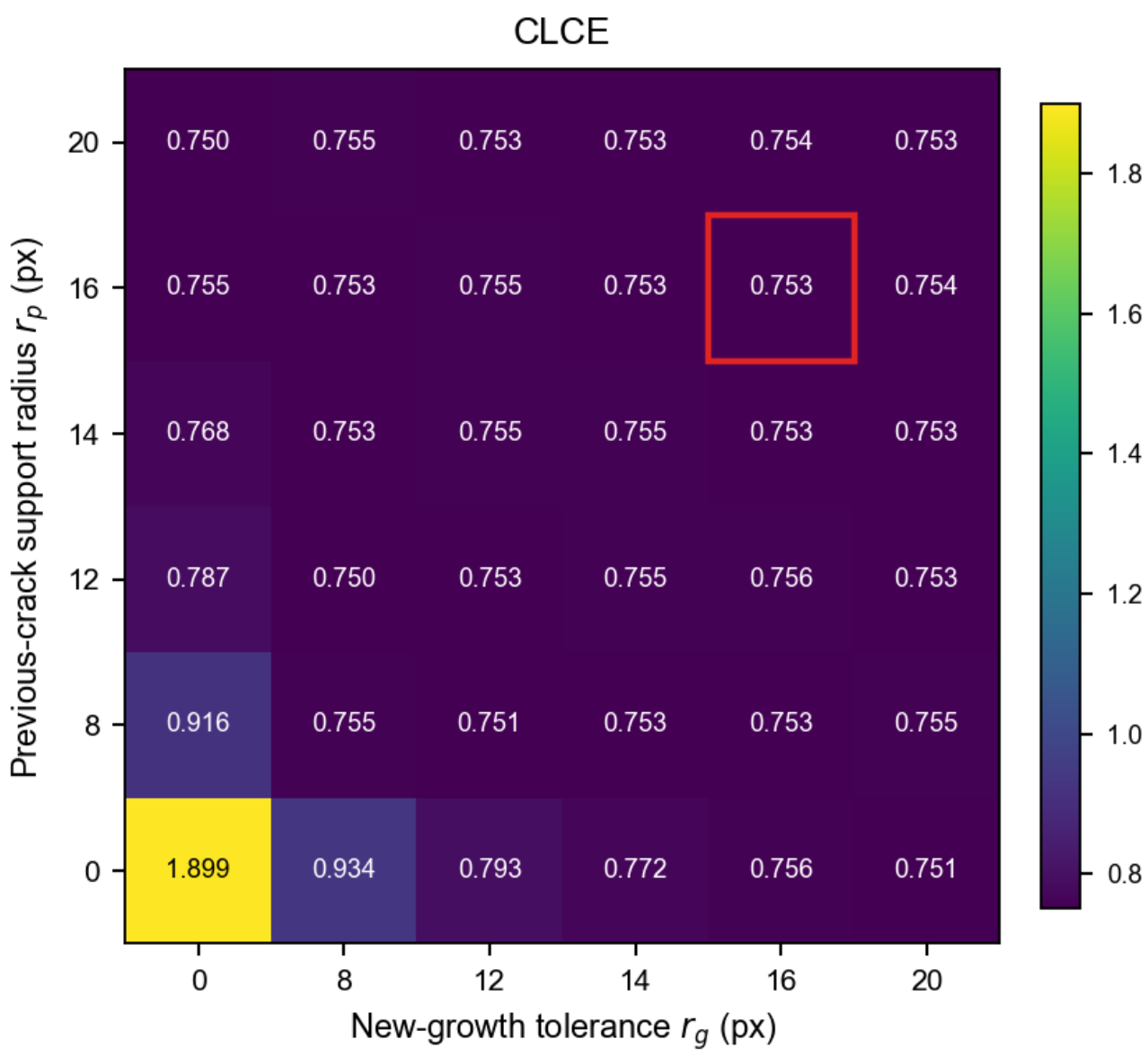

**Fig. 13.** Sensitivity of CLCE to the growth-support radii.

### 4.5 Error propagation and practical limitations

The dominant errors arise from a cascading interaction among incomplete early-crack segmentation, residual geometric projection error, and ambiguity in crack-instance formation and association. Nascent cracks are often only one to a few pixels wide and weakly contrasted; missing or discontinuous mask support may therefore fragment one physical crack, interrupt identity propagation, or delay the appearance of short branches. Although morphology-aware instantiation reconnects compatible fragments, it cannot recover crack support that is absent from the predicted mask. Residual projection drift under texture-poor, partially occluded, or strongly varying viewpoints further reduces overlap in the unified reference view. These uncertainties propagate into identity association, inherited-versus-new support separation, and growth-length estimation, contributing to lower HOTA under stricter overlap thresholds, scene-dependent CLCE, and delayed growth detection.

The evaluation is limited to eight laboratory scenes, a single fixed segmentation checkpoint, fixed pixel-scale tolerances, and multi-instance discrimination primarily in Scenes 6–8. No image from the evaluated temporal scenes was used to train or fine-tune this checkpoint; however, the sensitivity of the downstream results to alternative segmentation architectures or training data remains unexamined. Controlled staged loading isolates crack extension from viewpoint change, but field generalizability, scale transferability, and performance under irregular inspection intervals remain unverified. Reliable scale calibration is also required to convert image-space growth lengths to millimeters and measure crack centerlines along reconstructed three-dimensional surfaces. DGR@$K$ indicates whether reference crack growth is detected in its original interval or within the subsequent $K$ observation intervals, rather than quantifying the actual delay in physical time. The manually reviewed reference provides a consistent basis for comparison, although its construction relies on the quality of the available projection results. Evaluation independence could be further strengthened using calibrated control information or blinded reference construction. Future work will prioritize field validation, fine-crack segmentation, scale- and confidence-aware alignment, and calibrated, uncertainty-aware growth measurement.

## 5. Conclusion

This study innovatively formulates visual crack-growth monitoring under changing viewpoints as a cross-view and cross-time crack-instance tracking problem. To address this challenge, a unified reference-view framework is developed by integrating geometry-assisted alignment, morphology-aware crack instantiation, identity propagation, and pairwise growth attribution. The framework converts frame-wise crack masks into identity-consistent crack histories and distinguishes pre-existing crack regions from newly developed growth without requiring a fixed camera configuration.

A temporally annotated crack-image dataset was established to encompass representative cracking patterns, including Mode I opening cracks in pre-notched beams, Mode II-dominated diagonal shear cracks in deep beams, and mixed-mode flexural–shear cracks characterized by turning and branching in L-shaped beams. A dedicated evaluation protocol was further developed for quantitative assessment of crack-growth tracking. In addition to HOTA, which primarily evaluates cross-frame detection and identity association rather than the progressive growth, branching, and merging characteristics of cracks, a new CLCE metric was proposed to quantify post-association errors in crack-length growth, providing a more physically meaningful measure for thin, elongated, irregular, and topologically complex crack structures. Two auxiliary diagnostic metrics were also introduced to assess same-interval instance-level growth localization and the detection of reference growth within a specified observation window, respectively.

Experimental results demonstrate the feasibility of the proposed framework. The relatively low HOTA values highlight the limitations of conventional object-tracking metrics for crack tracking, as crack growth, branching, merging, and highly elongated morphology differ fundamentally from the motion and deformation of conventional object instances, although acceptable instance correspondence can still be achieved under a relaxed overlap criterion. Across the eight evaluated scenes, the framework achieved a pooled instance-level CLCE of 0.753, demonstrating the framework's capability to quantitatively characterize crack-length evolution. Overall, these results indicate that the proposed framework extends fixed-camera crack-growth detection to identity-aware crack-growth monitoring using overlapping images acquired from arbitrary viewpoints.

The proposed framework advances repeated visual crack inspection from isolated detections to identity-aware and quantitative crack-evolution monitoring under changing viewpoints. With further field validation, it could support automated condition assessment and maintenance prioritization. Current limitations include fine-crack segmentation errors, residual registration drift, fixed pixel-scale tolerances, and laboratory-scale validation. Future work will focus on field datasets, improved fine-crack extraction, registration-quality assessment, adaptive physical-scale criteria, and uncertainty-aware growth quantification.

## CRediT authorship contribution statement

Pengru Deng: Writing – review & editing, Project administration, Methodology, Funding acquisition, Conceptualization. Jiapeng Yao: Writing – original draft, Data curation, Validation, Formal analysis. Su Wang: Writing – review & editing, Project administration, Methodology. Xuhui He: Writing – review & editing, Funding acquisition, Supervision. Zhao Wang: Writing – review & editing, Project administration. Takanori Emaru: Writing – review & editing, Supervision. Takashi Matsumoto: Writing – review & editing, Supervision. Xinrun Li: Writing – original draft, Writing – review & editing, Methodology, Formal analysis.

## Declaration of interests

The authors declare that they have no known competing financial interests or personal relationships that could have appeared to influence the work reported in this paper.

## Data availability

Data will be made available on request.

## Acknowledgments

This work was supported by the National Natural Science Foundation of China (project No. 52308225, 2022HWYQ04), the Natural Science Foundation of Hunan Province, China (project No. 2025JJ90165, 2023JJ40721). Their support is gratefully acknowledged.